\documentclass[11pt, a4paper]{article}
\usepackage[a4paper, margin=2.5cm]{geometry}
\usepackage[T1]{fontenc}
\usepackage[utf8]{inputenc}
\usepackage{lmodern}

\usepackage{amsmath, amssymb, amsthm}
\usepackage{booktabs}
\usepackage{tabularx}
\usepackage{longtable}
\usepackage{microtype}
\usepackage{enumitem}
\usepackage[font=small,labelfont=bf,labelsep=period]{caption}
\usepackage{xcolor}
\usepackage{hyperref}

\definecolor{accentblue}{RGB}{24, 78, 138}
\definecolor{slategray}{RGB}{71, 85, 105}
\setlist[itemize,enumerate]{topsep=4pt plus 1pt, itemsep=2pt, parsep=0pt}
\newcommand{\besidetext}[3]{%
    \par\addvspace{5pt}%
    \noindent
    \begin{minipage}[t]{\dimexpr\linewidth-#1-1.2em\relax}%
        \vspace{0pt}\noindent\ignorespaces#3\par
    \end{minipage}\hfill
    \begin{minipage}[t]{#1}%
        \vspace{0pt}\small\noindent\ignorespaces#2\par
    \end{minipage}\par
    \addvspace{5pt}%
}
\hypersetup{
    pdftitle={Min-Plus Products and Dynamic Programming with Few Violations of the Adjacent Monge Inequalities},
    pdfauthor={Yanqiao Chen},
    colorlinks=true,
    linkcolor=accentblue,
    citecolor=accentblue,
    urlcolor=accentblue
}

\theoremstyle{plain}
\newtheorem{theorem}{Theorem}[section]
\newtheorem{lemma}[theorem]{Lemma}
\newtheorem{corollary}[theorem]{Corollary}
\newtheorem{proposition}[theorem]{Proposition}

\theoremstyle{definition}
\newtheorem{definition}[theorem]{Definition}
\newtheorem{example}[theorem]{Example}

\theoremstyle{remark}
\newtheorem{remark}[theorem]{Remark}

\newcounter{algorithm}
\newcommand{\algheading}[1]{%
    \refstepcounter{algorithm}\label{alg:#1}\textbf{Algorithm~\thealgorithm}%
}

\newenvironment{algorithmsteps}[2]{%
    \par\small
    \setlength{\LTpre}{6pt plus 2pt minus 2pt}%
    \setlength{\LTpost}{5pt plus 2pt minus 2pt}%
    \setlength{\LTleft}{0pt}%
    \setlength{\LTright}{0pt}%
    \begin{longtable}{@{}>{\raggedleft\arraybackslash}p{1.8em}@{\hspace{.7em}}>{\raggedright\arraybackslash}p{\dimexpr\linewidth-2.5em\relax}@{}}
    \toprule
    \multicolumn{2}{@{}l}{\algheading{#1}\quad #2}\\*
    \midrule
    \endfirsthead
    \toprule
    \multicolumn{2}{@{}l}{\textbf{Algorithm~\ref{alg:#1}}\quad #2 (continued)}\\*
    \midrule
    \endhead
}{%
    \bottomrule
    \end{longtable}%
}

\title{Min-Plus Products and Dynamic Programming\\with Few Violations of the Adjacent Monge Inequalities}
\author{Yanqiao Chen\thanks{Southern University of Science and Technology, Shenzhen, China. Email: \texttt{12412115@mail.sustech.edu.cn}}}
\date{}

\begin{document}

\maketitle

\begin{abstract}
Local rewards can introduce negative adjacent mixed differences into a convex transition cost. If a cost array $B$ has $s(B)$ such differences, we express it as a Monge array minus $s(B)$ weighted rectangles and sweep its columns, maintaining the lower envelope of the active rows. For explicit finite $n\times n$ matrices $A,B$ and $L=\lceil\log_2(n+1)\rceil$, we compute $A\otimes B$ and the smallest minimizing index for every entry in $\mathcal{O}(n^2L+ns(B)L^3)$ time and $\mathcal{O}(n^2)$ space, giving near-quadratic time for arbitrary $A$ when $s(B)=\mathcal{O}(n)$, including right factors with $\Theta(n^2)$ positive differences.

For block costs given by a constant-time Monge oracle and $r$ signed rectangle corrections, the sweep activates each row when its prefix optimum becomes available and computes an optimal contiguous partition in $\mathcal{O}((n+r)L^3)$ time and $\mathcal{O}(n+r)$ working space. The sweep handles convex transitions with interval rules and assigns one rectangle to each internal edge or hyperedge reward in a convex block cost, giving near-linear exact algorithms on sparse inputs for ordered graph modularity, weighted contiguous cluster editing, all-or-nothing HyperLam, and bounded-rank strict hypergraph modularity.

Let $c_{\mathrm{APSP}}$ denote the deterministic exponent of all-pairs shortest paths with polynomially bounded integer weights. For fixed $3-c_{\mathrm{APSP}}<\alpha\le1$, any deterministic $\mathcal{O}(n^\beta)$-time min-plus product algorithm for polynomially bounded integer matrices with $s(B)\le n^{1+\alpha}$ satisfies $\beta\ge c_{\mathrm{APSP}}-1+\alpha$. The reduction from general integer min-plus multiplication uses right factors obtained from $(j-k)^2$ by at most $n^{1+\alpha}$ interval discounts, with $\Theta(n^2)$ positive differences and all remaining differences negative.
\end{abstract}

\section{Introduction}

In min-plus multiplication and sequential dynamic programming, each output is a minimum over candidate rows whose costs include a row-dependent offset. When the transition matrix is Monge, SMAWK~\cite{Aggarwal1987} computes the product of an arbitrary $n\times n$ matrix with that transition matrix in $\mathcal{O}(n^2)$ time, and the algorithms of Wilber and of Larmore and Schieber~\cite{Wilber1988, Larmore1991} evaluate a triangular Monge recurrence in linear time. We study how these running times depend on the number of adjacent Monge inequalities violated by local rewards or interval adjustments to the transition costs.

For fixed row and column orders, define the adjacent mixed difference of a cost array $C$ by
\[
\Delta C_{p,q} = C(p+1,q) + C(p,q+1) - C(p,q) - C(p+1,q+1),
\]
and measure its negative support by
\[
\mathcal{N}(C) = \{(p,q) : \Delta C_{p,q} < 0\}, \qquad s(C) = |\mathcal{N}(C)|.
\]
The adjacent Monge inequalities require $\Delta C_{p,q}\ge0$ at every cell, and algorithms for core-sparse Monge multiplication~\cite{Tiskin2015, Russo2010, Gawrychowski2025} use the number $\delta(C)$ of nonzero adjacent differences. The parameter $s(C)$ counts the cells that violate the inequality, so an array with small $s(C)$ can have $\Theta(n^2)$ positive differences. Adding row and column potentials $a_k+b_j$ preserves every mixed difference and hence $s(C)$, including when the row potentials are entries $A_{ik}$ of a fixed row of the left factor or prefix optima $D[k]$ in a recurrence.

\subsection{Main results}

We write $L=\lceil\log_2(n+1)\rceil$ and use the same tree of row minima for matrix products, where the offsets are supplied with each input row, and for sequential recurrences, where the computation supplies an offset when it reaches the corresponding state. Table~\ref{tab:main-summary} gives the bounds for each input representation.

\begin{table}[!htbp]
\centering
\caption{Time and working space for products and partitioning, with $L=\lceil\log_2(n+1)\rceil$ and constant-time entry or penalty oracles. Space for convex transitions is given per vector. For $s(B)\le n^{1+\alpha}$ and fixed $3-c_{\mathrm{APSP}}<\alpha\le1$, the product lower exponent is $c_{\mathrm{APSP}}-1+\alpha$.}
\label{tab:main-summary}
\small
\setlength{\tabcolsep}{4pt}
\begin{tabularx}{\linewidth}{@{}>{\raggedright\arraybackslash}p{.225\linewidth} >{\raggedright\arraybackslash}X l l@{}}
\toprule
\textbf{Problem} & \textbf{Input} & \textbf{Time} & \textbf{Space} \\
\midrule
Matrix product (Thm.~\ref{thm:explicit-product}) & Explicit $A,B$ with $s=s(B)$ & $\mathcal{O}(n^2L+nsL^3)$ & $\mathcal{O}(n^2)$ \\
Vector product (Thm.~\ref{thm:static-transfer}) & Monge oracle and $r$ rectangles & $\mathcal{O}(nL+rL^3)$ & $\mathcal{O}(n+r)$ \\
Partition DP (Thm.~\ref{thm:sequential-dp}) & Monge oracle and $r$ rectangles & $\mathcal{O}((n+r)L^3)$ & $\mathcal{O}(n+r)$ \\
Partition DP (Cor.~\ref{cor:neg-list-dp}) & Cost oracle and list of $s$ negative cells & $\mathcal{O}((n+s)L^3)$ & $\mathcal{O}(n+sL)$ \\
Convex transitions (Cor.~\ref{cor:convex-rules}) & Coordinates, convex penalty, $r$ rectangles & \shortstack[l]{$\mathcal{O}(nL+rL^3)$ per vector\\$\mathcal{O}(n^2L+nrL^3)$ per matrix} & $\mathcal{O}(n+r)$ \\
Partition with rewards (Thm.~\ref{thm:interaction-dp}) & Vertex masses and $h$ interactions of incidence size $N$ & $\mathcal{O}(N+(n+h)L^3)$ & $\mathcal{O}(n+h)$ \\
\bottomrule
\end{tabularx}
\end{table}

\paragraph{Min-plus matrix multiplication.}
For explicit finite $n\times n$ matrices $A,B$ with $s=s(B)$, Theorem~\ref{thm:explicit-product} computes $A\otimes B$ and the smallest minimizing index for each entry in $\mathcal{O}(n^2L+nsL^3)$ time and $\mathcal{O}(n^2)$ space, which gives $\mathcal{O}(n^2L^3)$ time when $s=\mathcal{O}(n)$ and allows every other cell of $\Delta B$ to be strictly positive.

\paragraph{Sequential partitioning.}
Given a Monge background oracle $M$ with query time $T_M$ and $r$ rectangles $I_\ell\times J_\ell$ with signed weights $\alpha_\ell\in\mathbb{R}$, Theorem~\ref{thm:sequential-dp} evaluates the recurrence
\[
D[j] = \min_{0 \le k < j} \left\{ D[k] + M(k,j) + \sum_{\ell=1}^{r} \alpha_{\ell} \mathbf{1}\{k \in I_{\ell}\}\mathbf{1}\{j \in J_{\ell}\} \right\}
\]
and reconstructs an optimal partition in $\mathcal{O}((n+r)L^2(L+T_M))$ time and $\mathcal{O}(n+r)$ working space. Given constant-time access to $C$ and the complete list of its $s$ negative legal differences, a persistent segment tree evaluates $M$ in $\mathcal{O}(L)$ time per query, giving $\mathcal{O}((n+s)L^3)$ total time and $\mathcal{O}(n+sL)$ additional space (Corollary~\ref{cor:neg-list-dp}). Applying the sweep to each layer of the recurrence for exactly $K$ blocks takes $\mathcal{O}(K(n+r)L^2(L+T_M))$ time (Corollary~\ref{cor:k-blocks}).

\paragraph{Convex transitions and contiguous partitions.}
For a convex transition cost $\phi(y_j-x_k)$ with $r$ interval surcharges or discounts, $s(B)\le2r$ and the list of rules gives an $\mathcal{O}(nL+rL^3)$-time vector product (Corollary~\ref{cor:convex-rules}). For a contiguous block with convex load cost $\phi(S_j-S_k)$ and internal rewards $\sum_{e\subseteq(k,j]}w_e$, each of the $h$ edges or hyperedges supplies one rectangle, so processing the incidence input of size $N$ and running the sweep takes $\mathcal{O}(N+(n+h)L^3)$ time for a constant-time penalty. Section~\ref{sec:applications} gives the resulting algorithms for ordered graph modularity~\cite{Patania2023}, weighted contiguous cluster editing, all-or-nothing HyperLam~\cite{Veldt2020}, bounded-rank strict hypergraph modularity~\cite{Kaminski2019}, and FENNEL with a fixed number of blocks~\cite{Tsourakakis2014}.

\paragraph{Product lower bounds.}
For fixed $0<\alpha\le1$, Theorem~\ref{thm:fixed-quad-lb} reduces general min-plus multiplication on polynomially bounded integer entries to $\mathcal{O}(n^{1-\alpha})$ products with $s(B)\le n^{1+\alpha}$, so a deterministic $\mathcal{O}(n^\beta)$-time algorithm for these products gives an $\mathcal{O}(n^{1-\alpha+\beta}+n^{3-\alpha})$-time algorithm for the general problem. For $3-c_{\mathrm{APSP}}<\alpha\le1$, the resulting lower exponent $c_{\mathrm{APSP}}-1+\alpha$ and the sweep's upper exponent $2+\alpha$ differ by $3-c_{\mathrm{APSP}}$. The right factors used in the reduction have background $(j-k)^2$, at most $n^{1+\alpha}$ interval discounts, $\Theta(n^2)$ positive mixed differences, and a negative difference at every remaining cell (Section~\ref{sec:lower-bounds}).

\subsection{The column sweep}

\paragraph{Rectangle corrections.}
The indicator $Q_{p,q}(k,j)=\mathbf{1}\{k\le p\}\mathbf{1}\{j>q\}$ has adjacent mixed difference $1$ at $(p,q)$ and $0$ at every other cell.

\besidetext{.30\linewidth}{%
    \centering
    \begin{tabular}{@{}c|cc@{}}
        $Q_{p,q}$ & $j\le q$ & $j>q$ \\
        \hline
        $k\le p$ & $0$ & $1$ \\
        $k>p$ & $0$ & $0$
    \end{tabular}\par
}{%
    By linearity, adding $R(k,j)=\sum_{(p,q)\in\mathcal{N}(C)}(-\Delta C_{p,q})Q_{p,q}(k,j)$ to $C$ gives $\Delta(C+R)_{p,q}=\max\{\Delta C_{p,q},0\}$, so telescoping shows that $M=C+R$ is Monge and $C=M-R$ uses $s(C)$ rectangles, each consisting of a row prefix and a column suffix.
}
For a supplied rectangle $I_\ell\times J_\ell$ of weight $\alpha_\ell$, the sweep adds $\alpha_\ell$ on $I_\ell$ at column $\min J_\ell$ and subtracts it at $\max J_\ell+1$ if that column is in the domain, giving at most $2r$ interval additions for $r$ rectangles.

\paragraph{Lower envelopes of rows.}
At column $j$ an active row $k$ defines a function $f_k(x)=v_k+M(k,x)$ for $x\ge j$, where $v_k$ is the base potential plus the weights of the rectangles currently covering the row. For two nonempty sets of active rows $A,B$ with every row of $A$ preceding every row of $B$, the difference of their lower envelopes $F_A(x)-F_B(x)$ is nondecreasing by Lemma~\ref{lem:ordered-envelopes}, so the columns where $F_B(x)<F_A(x)$ form a possibly empty suffix whose first column determines the switch between the two sets.

Our balanced binary tree $T$ on $[0,n-1]$ follows Russo's switching-point trees~\cite{Russo2010}, with each internal node $u$ storing the switch column $\theta_u$ of its children and a lazy addition $z_u$. Adding $\beta$ to all active rows in a subtree shifts both child envelopes by $\beta$ and preserves all switches in the subtree, so $\operatorname{ADD}(I,\beta)$ stores lazy additions at $\mathcal{O}(L)$ fully covered subtrees and uses binary search to rebuild switches at $\mathcal{O}(L)$ partially covered ancestors in $\mathcal{O}(L^2(L+T_M))$ total time.

\paragraph{Row activation.}
For a vector product, all row potentials $x_k$ are available at column $0$, so the tree activates the leaves together and constructs the switches from the leaves to the root in $\mathcal{O}(nL(1+T_M))$ time. In a sequential recurrence the sweep activates row $k=j-1$ after computing $D[k]$, using a Fenwick tree $H$ that records the total weight $H_j(k)=\sum_\ell\alpha_\ell\mathbf{1}\{k\in I_\ell\}\mathbf{1}\{j\in J_\ell\}$ for every row, including rectangles whose entry event preceded activation. Each event updates $H$ in $\mathcal{O}(L)$ time and updates the envelope tree on the part of its row interval in $[0,j-2]$, after which a point query in $H$ gives the new row its potential $D[k]+H_j(k)$.

\paragraph{Interval rewards and the product reduction.}
An edge or hyperedge $e$ belongs to the block $(k,j]$ precisely when $k<\min e$ and $j\ge\max e$, so a scan of its $|e|$ vertices gives the rectangle $[0,\min e-1]\times[\max e,n]$ of weight $-w_e$ used with the convex block cost $\phi(S_j-S_k)$. In the product reduction of Theorem~\ref{thm:fixed-quad-lb}, scaling the integer costs by $\lambda=2n^2+1$ and adding a squared-distance background gives an adjusted cost $\lambda t-2kj$ for a candidate of original cost $t$, where $0\le2kj<\lambda$, so dividing the adjusted minimum by $\lambda$ and rounding up recovers the original entry.

\subsection{Related work}

\paragraph{Monge matrix products.}
SMAWK~\cite{Aggarwal1987} finds the row minima of a totally monotone $n\times n$ matrix in $\mathcal{O}(n)$ evaluations, and Tiskin~\cite{Tiskin2015} gave an $\mathcal{O}(n\log n)$ multiplication algorithm for simple unit-Monge matrices. For general Monge factors, Russo~\cite{Russo2010, Russo2012} developed switching-point trees whose product construction depends on the core size of the left factor, and Gawrychowski, Gorbachev, and Kociumaka~\cite{Gawrychowski2025} compute the condensed product of two condensed Monge matrices in $\mathcal{O}(n+\delta\log(1+\delta))$ time for total input core size $\delta$. Appendix~\ref{app:russo-alt} derives the time bound of Theorem~\ref{thm:explicit-product} for product values by encoding the negative differences of the right factor in an auxiliary Monge matrix and querying an antidiagonal of Russo's implicit product.

\paragraph{Monge recurrences and convolutions.}
In online algorithms for Monge recurrences and shortest paths~\cite{Wilber1988, Larmore1991, Gawrychowski2018}, rows become available sequentially and retain their assigned potentials, while the data structure of Charalampopoulos et al.~\cite{Charalampopoulos2022} supports sub-row increments in $\mathcal{O}(\log^3 n)$ time per update when each intermediate matrix is Monge. In our sweep an interval addition shifts the selected rows uniformly over future columns and activation supplies the next row potential, so at each step the active functions $v_k+M(k,x)$ form a Monge array, and changes in the row offsets from one step to the next can give the complete transition matrix $C$ negative differences.

For displacement costs $g(j-k)$, Felzenszwalb and Huttenlocher~\cite{Felzenszwalb2012} give linear-time convex distance transforms, and applying the convex-kernel convolution bound of Brand and Lassota~\cite{Brand2023} to the convex pieces of a kernel with $t$ negative second differences takes $\mathcal{O}(n(t+1))$ time. Our rectangle representation allows independent row and column intervals for each correction to a convex coordinate cost. Carmel, Tsur, and Ziv-Ukelson~\cite{Carmel2016, Carmel2019} bound the number of nonzero density entries of either sign in the all-scores matrix of an alignment grid with $b$ bridges by $\mathcal{O}(n(b+1))$.

\paragraph{Contiguous partitioning and APSP.}
Patania, Allard, and Young~\cite{Patania2023} give exact algorithms for graph modularity in $\mathcal{O}(n^2)$ time for a linear order and $\mathcal{O}(n^3)$ time for a cyclic order (see also Li, Cao, and Zhang~\cite{Li2020}), and Ahrens~\cite{Ahrens2021} obtains near-linear algorithms for contiguous communication objectives under quadrangle monotonicity. The block costs for modularity~\cite{Newman2004}, HyperLam~\cite{Veldt2020}, strict hypergraph modularity~\cite{Kaminski2019}, FENNEL~\cite{Tsourakakis2014}, and ordered cluster editing~\cite{Veldt2018, Bellitto2015, Lin2015} admit a convex penalty and internal rewards, which can give mixed differences of both signs.

Vassilevska Williams and Williams~\cite{Williams2018} established subcubic equivalences between APSP, min-plus products, and triangle problems, and Heeger, Nichterlein, and Niedermeier~\cite{Heeger2023} studied parameterized cross-compositions for Negative Triangle. For polynomially bounded integer weights, Alman and Vassilevska Williams~\cite{Alman2026} gave a deterministic $\mathcal{O}(n^{2.99942})$ algorithm for APSP and min-plus multiplication, and Saha and Ye~\cite{Saha2026} improved the multiplication bound to $\tilde{\mathcal{O}}(n^{2.9964})$.

\section{Preliminaries}
\label{sec:preliminaries}

We index square matrices by $[0,n-1]$ and partitions of $n$ ordered vertices by cuts $0,\ldots,n$, with $(k,j]$ denoting the vertices after cut $k$ and up to cut $j$. All intervals are integer intervals, with $[a,b]$ empty for $a>b$, and we write $\mathbf{1}\{P\}$ for the indicator of a predicate $P$ and $L=\lceil\log_2(n+1)\rceil$ as in Table~\ref{tab:notation}.

\paragraph{Monge arrays and mixed differences.}
On the square $[0,n-1]^2$ or the triangular domain $\mathcal{T}_n=\{(k,j):0\le k<j\le n\}$, a cost array $M$ is \emph{Monge} if
\begin{equation}
\label{eq:monge-def}
M(a,x) + M(b,y) \le M(a,y) + M(b,x)
\end{equation}
for all rows $a<b$ and columns $x<y$ for which the four entries are defined, a condition that requires $b<x$ on $\mathcal{T}_n$. A cell $(p,q)$ is \emph{legal} when all four entries of $\Delta C_{p,q}$ belong to the domain, which on $\mathcal{T}_n$ means $p\ge0$, $p+1<q$ and $q+1\le n$. Summing these differences over a rectangle gives
\begin{equation}
\label{eq:telescoping}
\sum_{p=a}^{b-1} \sum_{q=x}^{y-1} \Delta C_{p,q} = C(b,x) + C(a,y) - C(a,x) - C(b,y),
\end{equation}
so $C$ is Monge if and only if $\Delta C_{p,q} \ge 0$ on every legal cell. Moreover, for arbitrary row and column potentials $a_{k}, b_{j}$,
\begin{equation}
\label{eq:potential-invariance}
\Delta\bigl(C(k,j) + a_{k} + b_{j}\bigr)_{p,q} = \Delta C_{p,q}.
\end{equation}

\besidetext{.45\linewidth}{%
    \captionsetup{justification=raggedright,singlelinecheck=false,hypcap=false}
    \captionof{table}{Notation}\label{tab:notation}
    \setlength{\tabcolsep}{3pt}
    \begin{tabularx}{\linewidth}{@{}l >{\raggedright\arraybackslash}X@{}}
        \toprule
        $n,L$ & Dimension, $\lceil\log_2(n+1)\rceil$ \\
        $\mathcal{T}_n$ & Triangular domain \\
        $\Delta C_{p,q}$ & Adjacent mixed difference \\
        $\mathcal{N}(C),s(C)$ & Negative cells and their count \\
        $r$ & Number of rectangles \\
        $M,T_M$ & Monge background and query time \\
        $A\otimes B$ & Min-plus product \\
        $U$ & Bound on absolute input weights \\
        $G,W,d_i$ & Graph, total weight, degree \\
        $h,N,\rho$ & Hyperedge count, incidence size, rank \\
        $G_-(B)$ & Graph of negative cells \\
        $c_{\mathrm{APSP}}$ & Deterministic integer APSP exponent \\
        \bottomrule
    \end{tabularx}\par
}{%
    We count exact arithmetic operations and comparisons at unit cost and charge $T_M$ for an entry query to a supplied Monge background $M$. For polynomially bounded integer inputs and fixed-degree polynomial costs with polynomially bounded integer coefficients, the intermediate values fit in $\mathcal{O}(\log n)$ bits and the same bounds hold on the word RAM. The input representation determines the preprocessing. A scan of an explicit matrix finds the negative cells, a rectangle list supplies the row updates for a given Monge background, and a list of negative cells together with an oracle for the full cost supplies the data for constructing that background.
}

\paragraph{Matrix products and contiguous partitions.}
For finite $n \times n$ matrices $A, B$ indexed by $[0, n-1]$, their min-plus product is $(A \otimes B)_{ij} = \min_{0 \le k < n}\{A_{ik} + B_{kj}\}$, and a \emph{witness} is a minimizing index $k$. For contiguous partitioning of $[1,n]$, a partition is given by cut positions $0 = b_{0} < b_{1} < \dots < b_{K} = n$ with total cost $\sum_{i=1}^{K} C(b_{i-1}, b_{i})$. With an unrestricted number of blocks, the minimum cost $D[n]$ satisfies
\begin{equation}
\label{eq:dp-unrestricted}
D[0] = 0, \qquad D[j] = \min_{0 \le k < j} \{D[k] + C(k,j)\} \quad (1 \le j \le n),
\end{equation}
with optimal cuts reconstructed from $\pi[j]$ (breaking ties toward the smallest cut). With an exact block count $K$, the layered recurrence is
\begin{equation}
\label{eq:dp-k-blocks}
D_{b}[j] = \min_{0 \le k < j} \{D_{b-1}[k] + C(k,j)\} \quad (1 \le b \le K,\; 1 \le j \le n),
\end{equation}
initialized by $D_{0}[0] = 0$, $D_{0}[j] = +\infty$ for $j > 0$, and $D_{b}[0] = +\infty$ for $b > 0$.

\paragraph{The APSP exponent.}
The product lower bounds are expressed in terms of the deterministic exponent for APSP on polynomially bounded integer weights, for which truly subcubic algorithms are known~\cite{Alman2026, Saha2026}.

\begin{definition}[APSP exponent and $\gamma$-APSP hypothesis]
\label{def:gamma-apsp}
Let $c_{\mathrm{APSP}}$ be the infimum of the exponents $c$ for which a deterministic algorithm solves APSP on $n$-vertex directed graphs with polynomially bounded integer weights in $\mathcal{O}(n^c)$ time, and for a constant $\gamma\in(2,3]$ let the \emph{$\gamma$-APSP hypothesis} assert $c_{\mathrm{APSP}}\ge\gamma$. The output size gives $c_{\mathrm{APSP}}\ge2$, and the algorithm of Saha and Ye~\cite{Saha2026} gives $c_{\mathrm{APSP}}\le2.9964$.
\end{definition}

The classical reductions between APSP and min-plus multiplication (see~\cite{Williams2018}) preserve the exponent up to polylogarithmic factors in running time, so the deterministic exponent of min-plus multiplication for polynomially bounded integer $n\times n$ matrices is also $c_{\mathrm{APSP}}$.

\section{Matrix products and recurrences with rectangle corrections}
\label{sec:engine}

The sweep represents a cost array by a Monge background and a list of rectangles, processes the rectangle endpoints in column order, and maintains the minimum over active rows in a tree that supports interval additions and row activation (Appendix~\ref{app:data-structure}).

\subsection{Decomposition by negative mixed differences}

\begin{lemma}[Decomposition by negative differences]
\label{lem:neg-decomp}
Let $C$ be a finite rectangular or triangular cost array with negative support $\mathcal{N}(C)$. For each $(p,q) \in \mathcal{N}(C)$, define $w_{p,q} = -\Delta C_{p,q} > 0$, and set
\begin{equation}
\label{eq:decomp}
R(k,j) = \sum_{(p,q) \in \mathcal{N}(C)} w_{p,q} \mathbf{1}\{k \le p\}\mathbf{1}\{j > q\}, \qquad M = C + R.
\end{equation}
Then $\Delta M_{p,q}=\max\{\Delta C_{p,q},0\}$ at every legal cell, so $M$ is Monge and $C=M-R$.
\end{lemma}

\begin{proof}
The indicator $Q_{p,q}(k,j)=\mathbf{1}\{k\le p\}\mathbf{1}\{j>q\}$ has mixed difference $0+1-0-0=1$ at $(p,q)$ and zero at all other legal cells, so linearity gives $\Delta M_{p,q}=\Delta C_{p,q}+\sum_{(p',q')\in\mathcal{N}(C)}w_{p',q'}\mathbf{1}\{(p',q')=(p,q)\}=\max\{\Delta C_{p,q},0\}$. On both domains the cells in each rectangle of~\eqref{eq:telescoping} are legal, so their nonnegative sum proves the Monge inequality for $M$.
\end{proof}

More generally, consider any cost array given in the signed-rectangle representation
\begin{equation}
\label{eq:rect-model}
C(k,j) = M(k,j) + \sum_{\ell=1}^{r} \alpha_{\ell} \mathbf{1}\{k \in I_{\ell}\}\mathbf{1}\{j \in J_{\ell}\},
\end{equation}
where $M$ is Monge, $I_{\ell}$ and $J_{\ell}$ are integer intervals, and $\alpha_{\ell} \in \mathbb{R}$ may have either sign. During a left-to-right column sweep $j$, a rectangle with $J_{\ell} = [a_{\ell}, b_{\ell}]$ generates an entry event $(I_{\ell}, +\alpha_{\ell})$ at column $a_{\ell}$ and an exit event $(I_{\ell}, -\alpha_{\ell})$ at column $b_{\ell} + 1$ (if $b_{\ell} + 1$ lies in the column domain). Bucketing these at most $2r$ events takes $\mathcal{O}(n+r)$ time and space. Lemma~\ref{lem:neg-decomp} corresponds to $r = s(C)$ prefix rectangles $[0,p] \times [q+1, \nu]$ with weights $-w_{p,q}$.

\subsection{Lower envelopes of active rows}

At sweep column $j$, an active row $k$ with effective potential $v_k$ defines the cost function $f_k(x)=v_k+M(k,x)$ for $j\le x\le\nu$, where the last column is $\nu=n-1$ for a square matrix and $\nu=n$ for a triangular recurrence.

\begin{lemma}[Lower envelopes of ordered row sets]
\label{lem:ordered-envelopes}
For nonempty sets $A,B$ of active rows with $a<b$ for every $a\in A$ and $b\in B$, let $F_A(x)=\min_{a\in A}f_a(x)$ and $F_B(x)=\min_{b\in B}f_b(x)$. Their difference $F_A(x)-F_B(x)$ is nondecreasing on $[j,\nu]$, so the set of columns on which $F_B(x)<F_A(x)$ is a suffix of this interval.
\end{lemma}

\begin{proof}
For $j\le x<y\le\nu$, choose $a\in A$ attaining $F_A(y)$ and $b\in B$ attaining $F_B(x)$. The four entries lie in the domain, with $a<b<j\le x<y$ in the triangular case, so the Monge inequality~\eqref{eq:monge-def} gives $f_a(y)-f_b(y)\ge f_a(x)-f_b(x)$ and therefore
\[
F_{A}(y) - F_{B}(y) \ge f_{a}(y) - f_{b}(y) \ge f_{a}(x) - f_{b}(x) \ge F_{A}(x) - F_{B}(x). \qedhere
\]
\end{proof}

Each node $u$ of the balanced binary tree $T$ on the row indices $[0,n-1]$ stores its row interval $K(u)$ and its number of active leaves, and each internal node stores a lazy uniform addition $z_u$ and a \emph{switch column} $\theta_u$. At column $j$, rebuilding a switch sets $\theta_u$ to the first $x\in[j,\nu]$ where the right child's envelope is strictly smaller, with $\theta_u=j$ when only the right child is nonempty and $\theta_u=\nu+1$ when the right child is empty or never strictly smaller. A tie goes to the left child, giving the smallest minimizing row index, and a switch recorded at an earlier step remains valid on the smaller domain of future columns, where a switch below $j$ sends every query to the right child.

\begin{lemma}[Operations on the envelope tree]
\label{lem:tree-bounds}
The tree $T$ takes $\mathcal{O}(n)$ space and $\mathcal{O}(n)$ time to initialize and supports the following operations at sweep column $j$.
\begin{enumerate}
    \item $\operatorname{EVAL}(T.\mathrm{root}, x)$ ($x \ge j$) returns $\min_{k \text{ active}} \{v_{k} + M(k,x)\}$ and its smallest minimizing row in $\mathcal{O}(L + T_{M})$ time.
    \item $\operatorname{ADD}(T.\mathrm{root}, I, \beta, j)$ adds $\beta$ to all active rows in $I$ in $\mathcal{O}(L^{2}(L + T_{M}))$ time.
    \item $\operatorname{ACTIVATE}(T.\mathrm{root}, k, v, j)$ activates leaf $k$ with potential $v$ in $\mathcal{O}(L^{2}(L + T_{M}))$ time.
    \item Querying the minimum at $x \ge j$ restricted to active rows in a contiguous interval $F_{j}$ takes $\mathcal{O}(L(L + T_{M}))$ time.
\end{enumerate}
\end{lemma}

\begin{proof}
An evaluation follows one root-to-leaf path according to $x<\theta_u$, adds the lazy tags on that path, and queries $M(k,x)$ once at the leaf, for a total cost of $\mathcal{O}(L+T_M)$. By Lemma~\ref{lem:ordered-envelopes}, $\operatorname{PULL}(u,j)$ finds the switch by binary search with endpoint $\nu+1$, using $\mathcal{O}(L)$ evaluations in $\mathcal{O}(L(L+T_M))$ time. For $\operatorname{ADD}(I,\beta)$, a fully covered active leaf receives $\beta$ directly and a fully covered nonempty internal node adds $\beta$ to its lazy tag, which shifts both child envelopes equally and preserves their switches. Rebuilding the switches of the $\mathcal{O}(L)$ partially covered ancestors takes $\mathcal{O}(L^2(L+T_M))$ time, as does activating a leaf and rebuilding its ancestors (Algorithms~\ref{alg:eval-pull} and~\ref{alg:add-activate}).

For a query restricted to $F_j$, the descent partitions $F_j$ into $\mathcal{O}(L)$ canonical subtrees and carries the sum of the lazy tags of their strict ancestors, so adding this sum to each subtree's evaluation and taking the lexicographically smallest pair of value and row index takes $\mathcal{O}(L(L+T_M))$ time. A query with no active row in $F_j$ returns $(+\infty,-1)$.
\end{proof}

\subsection{Vector and matrix products}

When the row potentials $x_0,\ldots,x_{n-1}$ are available at column $0$, the algorithm activates all leaves together and computes the switches from the leaves to the root.

\begin{theorem}[Vector products with rectangle corrections]
\label{thm:static-transfer}
Given a constant-time Monge oracle $M$ and $r$ signed rectangles defining an $n\times n$ matrix $C$ by~\eqref{eq:rect-model}, the sweep groups events by column in $\mathcal{O}(n+r)$ time and space, after which it computes $x\otimes C$ and the smallest minimizing index for each coordinate of any finite $x\in\mathbb{R}^n$ in $\mathcal{O}(nL+rL^3)$ time and $\mathcal{O}(n+r)$ working space.
\end{theorem}

\begin{proof}
Algorithm~\ref{alg:static-transfer} installs $x_k$ at leaf $k$ and runs $\operatorname{PULL}(u,0)$ from the leaves to the root, spending $\mathcal{O}((h+1)L)$ time at a node of height $h$ and $\mathcal{O}(nL\sum_{h\ge0}(h+1)/2^h)=\mathcal{O}(nL)$ time over the tree. At each column $j=0,\ldots,n-1$, the algorithm applies the events in bucket $j$ using $\operatorname{ADD}$, which gives row $k$ the potential $x_k+\sum_{\ell=1}^{r}\alpha_\ell\mathbf{1}\{k\in I_\ell\}\mathbf{1}\{j\in J_\ell\}$, so $\operatorname{EVAL}(T.\mathrm{root},j)$ returns $(x\otimes C)_j$ and its smallest witness. The sweep makes at most $2r$ updates in $\mathcal{O}(rL^3)$ time and $n$ queries in $\mathcal{O}(nL)$ time.
\end{proof}

\begin{theorem}[Matrix products with sparse negative differences]
\label{thm:explicit-product}
For explicit finite $n\times n$ matrices $A,B$ with arbitrary $A$ and $s=s(B)$, the product $A\otimes B$ and the smallest minimizing index for each entry can be computed in $\mathcal{O}(n^2L+nsL^3)$ time and $\mathcal{O}(n^2)$ space, giving $\mathcal{O}(n^2L^3)$ time when $s=\mathcal{O}(n)$ for any number of positive differences up to $\Theta(n^2)$.
\end{theorem}

\begin{proof}
A scan of $\Delta B$ identifies the $s$ cells in $\mathcal{N}(B)$, and two-dimensional prefix sums construct $R(k,j)=\sum_{(p,q)\in\mathcal{N}(B)}(-\Delta B_{p,q})\mathbf{1}\{k\le p\}\mathbf{1}\{j>q\}$ in $\mathcal{O}(n^2)$ time. By Lemma~\ref{lem:neg-decomp}, $M=B+R$ is Monge and the representation $B=M-R$ gives one event $([0,p],\Delta B_{p,q})$ at column $q+1$ for each negative cell, so Theorem~\ref{thm:static-transfer} computes the product row and its witnesses for $x=A_{i,*}$ in $\mathcal{O}(nL+sL^3)$ time, reusing the $\mathcal{O}(n)$ tree storage across rows (Algorithm~\ref{alg:explicit-product}). For $|B_{kj}|\le U$, the bound $0\le R(k,j)\le4(n-1)^2U$ gives word-sized background entries on polynomially bounded integer inputs.
\end{proof}

\paragraph{Adaptive queries with a fixed right factor.}
For a fixed explicit $B$, the $\mathcal{O}(n^2)$ preprocessing constructs the background and event buckets once, so a batch of $h$ finite vectors, each of which may depend on the answers to previous queries, takes
\[
\mathcal{O}\bigl(n^2+h(nL+s(B)L^3)\bigr)
\]
total time, since each query initializes its own envelope tree and returns the smallest minimizing index for every output coordinate. Processing the queries in sequence uses $\mathcal{O}(n^2)$ working space, and writing the output takes $\Omega(n)$ time for a vector and $\Omega(n^2)$ for a matrix product. Corollary~\ref{cor:adaptive-query-lower} bounds the amortized query time for quadratic preprocessing.

\subsection{Sequential partitioning}

In recurrence~\eqref{eq:dp-unrestricted}, row $k$ becomes active at column $k+1$ with the newly computed prefix value $D[k]$, and its initial potential must include the weights of rectangles already covering that row. At column $j$ the envelope tree applies each interval addition to the previously considered rows in $[0,j-2]$, and a Fenwick tree $H$ over a difference array on $[0,n-1]$ records the sum for every row,
\[
H_{j}(k) = \sum_{\ell=1}^{r} \alpha_{\ell}\mathbf{1}\{k \in I_{\ell}\}\mathbf{1}\{j \in J_{\ell}\}.
\]
Each event $(I, \beta)$ updates $H$ in $\mathcal{O}(L)$ time, and querying $H_{j}(k)$ takes $\mathcal{O}(L)$ time.

\begin{theorem}[Sequential partitioning with rectangle corrections]
\label{thm:sequential-dp}
Given~\eqref{eq:rect-model} on $\mathcal{T}_{n}$ with background query time $T_{M}$, the recurrence~\eqref{eq:dp-unrestricted} and an optimal partition can be computed in $\mathcal{O}((n+r)L^{2}(L+T_{M}))$ time and $\mathcal{O}(n+r)$ working space (excluding background oracle storage). The same bounds hold when each state $j$ restricts its predecessors to a supplied interval $F_{j} \subseteq [0, j-1]$.
\end{theorem}

\begin{proof}
Immediately before the query at column $j$, the active leaves are the reachable states $k<j$ and row $k$ has effective potential $D[k]+H_j(k)$. Processing bucket $j$ updates $H_{j-1}$ to $H_j$ and applies $\operatorname{ADD}(T.\mathrm{root},I\cap[0,j-2],\beta,j)$ to the active part of each affected interval, after which a finite $D[j-1]$ activates row $j-1$ with potential $D[j-1]+H_j(j-1)$. The minimum over $F_j$ therefore equals $\min_{k\in F_j}\{D[k]+H_j(k)+M(k,j)\}=\min_{k\in F_j}\{D[k]+C(k,j)\}$, which proves the recurrence by induction. By Lemma~\ref{lem:tree-bounds}, the at most $2r$ events, $n$ activations and $n$ queries take $\mathcal{O}((n+r)L^2(L+T_M))$ time (Algorithm~\ref{alg:seq-dp}), and the stored predecessors reconstruct an optimal partition whenever the final value is finite, with $+\infty$ denoting infeasibility.
\end{proof}

\begin{corollary}[Sequential partitioning from a list of negative differences]
\label{cor:neg-list-dp}
Given constant-time access to $C$ on $\mathcal{T}_{n}$ and the complete list of its $s$ negative legal adjacent differences, \eqref{eq:dp-unrestricted} can be solved in $\mathcal{O}((n+s)L^{3})$ time and $\mathcal{O}(n+sL)$ additional space.
\end{corollary}

\begin{proof}
Grouping negative cells by $q$ gives a persistent segment tree on row gaps $[0,n-1]$ whose version $V_j$ stores the weights $w_{p,q}=-\Delta C_{p,q}$ for $q<j$, so $M(k,j)=C(k,j)+\operatorname{SuffixSum}(V_j,k)$ takes $\mathcal{O}(L)$ time per query after $\mathcal{O}(n+sL)$ construction time and space (Algorithm~\ref{alg:persistent-oracle}). Applying Theorem~\ref{thm:sequential-dp} with $r=s$ gives the stated $\mathcal{O}((n+s)L^3)$ bound, and an explicit triangular array can supply the negative list by an initial $\mathcal{O}(n^2)$ scan.
\end{proof}

\begin{corollary}[Partitions into $K$ blocks]
\label{cor:k-blocks}
The layered recurrence~\eqref{eq:dp-k-blocks} for exactly $K$ blocks can be solved in $\mathcal{O}(K(n+r)L^{2}(L+T_{M}))$ time using $\mathcal{O}(n+r)$ working space with rolling arrays, plus $\mathcal{O}(Kn)$ space if all predecessor pointers are retained for reconstruction.
\end{corollary}

\section{Applications}
\label{sec:applications}

For convex state transitions the input specifies the row and column coordinates and a list of interval adjustments, and for contiguous partitions it specifies vertex masses and the edges or hyperedges rewarded inside each block. In both cases these data give the Monge background and the rectangle events directly.

\subsection{Convex transition costs}

For nondecreasing coordinates $x_0\le\cdots\le x_{n-1}$ and $y_0\le\cdots\le y_{n-1}$ and a convex function $\phi:\mathbb{R}\to\mathbb{R}$ evaluable in constant time, the transition matrix with $r$ interval surcharges or discounts has entries
\begin{equation}
\label{eq:convex-rules}
B_{kj} = \phi(y_{j} - x_{k}) + \sum_{\ell=1}^{r} \alpha_{\ell}\mathbf{1}\{k \in I_{\ell}\}\mathbf{1}\{j \in J_{\ell}\}.
\end{equation}

\begin{lemma}[Mixed differences of a rectangle]
\label{lem:convex-corners}
In~\eqref{eq:convex-rules}, the background $M(k,j)=\phi(y_j-x_k)$ is Monge and each signed rectangle contributes at most four nonzero adjacent mixed differences, at most two of which are negative, so $s(B)\le2r$.
\end{lemma}

\begin{proof}
For $p \in [0, n-2]$ and $q \in [0, n-2]$, put $z = y_{q} - x_{p+1}$, $u = x_{p+1} - x_{p} \ge 0$, and $v = y_{q+1} - y_{q} \ge 0$. Convexity implies that increments $\phi(t+u) - \phi(t)$ are nondecreasing in $t$, so $\Delta M_{p,q} = \phi(z) + \phi(z+u+v) - \phi(z+u) - \phi(z+v) \ge 0$. For a single rectangle $\alpha \mathbf{1}\{k \in I\}\mathbf{1}\{j \in J\}$, writing $a_{k} = \mathbf{1}\{k \in I\}$ and $b_{j} = \mathbf{1}\{j \in J\}$ gives
\[
\Delta(\alpha a_{k}b_{j})_{p,q} = -\alpha (a_{p+1} - a_{p})(b_{q+1} - b_{q}),
\]
which can be nonzero only at the four corner gaps where both indicators change, with at most two negative entries because the signs alternate, so the union of the negative cells of the $r$ rectangles has size at most $2r$ and contains every negative cell of $B$.
\end{proof}

For $\phi(z)=z^2$ the background difference is $2(x_{p+1}-x_p)(y_{q+1}-y_q)$, so with strictly increasing coordinates at least $(n-1)^2-4r$ positive differences remain after the $r$ rectangles modify at most $4r$ cells.

\begin{corollary}[Vector products and layered recurrences]
\label{cor:convex-rules}
Let $B$ be given by the coordinates, the constant-time oracle for $\phi$, and the $r$ rectangle rules in~\eqref{eq:convex-rules}.
\begin{enumerate}
    \item After $\mathcal{O}(n+r)$ preprocessing, the algorithm computes $v\otimes B$ and its witnesses for any finite vector $v$ in $\mathcal{O}(nL+rL^3)$ time and $\mathcal{O}(n+r)$ space, and using the same rules for every row of $A$ computes $A\otimes B$ in $\mathcal{O}(n^2L+nrL^3)$ time.
    \item For an $\ell$-layer chain $d_{t}(j) = u_{t}(j) + \min_{k}\{d_{t-1}(k) + (B_{t})_{kj}\}$ with arbitrary finite unary costs $u_{t}(j)$ and $r_{t}$ rules at layer $t$, computing $d_{\ell}$ and an optimal path takes $\mathcal{O}(\ell nL + L^{3}\sum_{t=1}^{\ell} r_{t})$ time.
\end{enumerate}
\end{corollary}

One chain uses $\mathcal{O}(n+\max_t r_t)$ working space beyond the supplied layers, with another $\mathcal{O}(\ell n)$ space for all predecessors if a path is to be recovered in $\mathcal{O}(\ell)$ time. Applying the transfer to each of $n$ initial vectors gives all source-to-terminal costs in $\mathcal{O}(\ell n^2L+nL^3\sum_t r_t)$ time from the same rules.

\paragraph{Toeplitz costs and alignment grids.}
For $B_{kj}=g(j-k)$, the mixed difference at $(p,q)$ is $g(z-1)+g(z+1)-2g(z)$ with $z=q-p$, so each of the $t$ negative second differences of $g$ contributes a diagonal of at most $n-1$ cells and $s(B)\le(n-1)t$. Splitting $g$ into at most $t+1$ convex pieces and taking the componentwise minimum of their convolutions~\cite{Brand2023} gives $\mathcal{O}(n(t+1))$ time per vector and $\mathcal{O}(n^2(t+1))$ time for a matrix product, while~\eqref{eq:convex-rules} allows arbitrary nondecreasing coordinates and separately specified row and column intervals.

For an $m\times n$ alignment grid with longest common subsequence scores and $b\ge1$ positively weighted bridges, Carmel, Tsur, and Ziv-Ukelson~\cite{Carmel2016} construct a finite all-scores matrix in $\mathcal{O}(b(nm+n^2))$ time by assigning score $j-i$ when $i>j$. This completion has $\mathcal{O}(n(b+1))$ nonzero density entries, so negating its scores and applying Theorem~\ref{thm:explicit-product} takes a further $\tilde{\mathcal{O}}(n^2(b+1))$ time.

\subsection{Contiguous partitions with convex costs and internal rewards}

Let $a_{1}, \dots, a_{n} \ge 0$ be vertex masses with prefix sums $S_{j} = \sum_{i=1}^{j} a_{i}$ ($S_{0} = 0$), let $\phi : \mathbb{R}_{\ge 0} \to \mathbb{R}$ be a convex load penalty evaluable in $T_{\phi}$ time, and let $\mathcal{E}$ be a list of $h$ nonempty interactions (edges or hyperedges) $e \subseteq [1,n]$ with weights $w_{e} \ge 0$ and total incidence size $N = \sum_{e \in \mathcal{E}} |e|$. Consider the contiguous block cost
\begin{equation}
\label{eq:interaction-cost}
C(k,j) = \phi(S_{j} - S_{k}) - \sum_{e \subseteq (k,j]} w_{e}.
\end{equation}

\begin{lemma}[The rectangle for an internal reward]
\label{lem:interaction-rect}
In~\eqref{eq:interaction-cost}, the background $M(k,j)=\phi(S_j-S_k)$ is Monge and each interaction $e\in\mathcal{E}$ contributes the rectangle $[0,\min e-1]\times[\max e,n]$ with weight $-w_e$, giving $s(C)\le h$.
\end{lemma}

\begin{proof}
For $0\le a<b<x<y\le n$, set $z=S_x-S_b\ge0$, $u=S_b-S_a\ge0$ and $v=S_y-S_x\ge0$, so convexity gives $\phi(z+u)+\phi(z+v)\le\phi(z)+\phi(z+u+v)$ and hence the Monge inequality for $M$. An interaction satisfies $e\subseteq(k,j]$ precisely when $k<\min e$ and $j\ge\max e$, which gives the stated rectangle, whose mixed difference is supported at $(\min e-1,\max e-1)$ if that cell is legal. Since the background has nonnegative differences, the negative support lies among these at most $h$ cells.
\end{proof}

\begin{theorem}[Contiguous partitions with internal rewards]
\label{thm:interaction-dp}
An optimal contiguous partition for~\eqref{eq:interaction-cost} can be computed in $\mathcal{O}(N+n+(n+h)L^2(L+T_\phi))$ time and $\mathcal{O}(n+h)$ working space beyond the incidence input, and an optimal partition with exactly $K$ blocks takes $\mathcal{O}(N+n+K(n+h)L^2(L+T_\phi))$ time. For $T_\phi=\mathcal{O}(1)$ the respective time bounds are $\mathcal{O}(N+(n+h)L^3)$ and $\mathcal{O}(N+K(n+h)L^3)$.
\end{theorem}

\begin{proof}
Computing $S_{j}$ and all endpoint pairs $(\min e, \max e)$ takes $\mathcal{O}(N + n)$ time. Lemma~\ref{lem:interaction-rect} supplies a Monge oracle with $T_{M} = T_{\phi}$ and $h$ rectangles, so Theorem~\ref{thm:sequential-dp} and Corollary~\ref{cor:k-blocks} apply directly (Algorithm~\ref{alg:interaction-dp}).
\end{proof}

A direct dynamic program groups interactions by their right endpoint and, at column $j$, adds the weights of the new interactions at their left endpoints before scanning the cuts backward to compute the internal rewards by suffix sums. This takes $\mathcal{O}(N+n^2T_\phi)$ time over all columns, or $\mathcal{O}(N+Kn^2)$ time for $K$ layers if the penalty takes constant time and the block costs are stored once. A fixed charge per block amounts to adding a constant to $\phi$.

\begin{table}[!htbp]
\centering
\caption{Running times for contiguous partitions in a fixed vertex order, with $m$ graph edges, $h$ hyperedges of total incidence size $N$, and $q$ listed pairs. The strict modularity bound assumes bounded rank, and FENNEL uses a constant-time exact penalty oracle.}
\label{tab:partition-apps}
\small
\setlength{\tabcolsep}{4pt}
\begin{tabularx}{\linewidth}{@{}>{\raggedright\arraybackslash}p{.225\linewidth} >{\raggedright\arraybackslash}X l l@{}}
\toprule
\textbf{Objective} & \textbf{Block cost or rule} & \textbf{Prior or direct DP} & \textbf{Running time} \\
\midrule
Ordered graph modularity & $a_{i} = d_{i}$, $\phi(x) = x^{2}$, $w_{e} = 4Wb_{e}$ & $\mathcal{O}(n^{2})$~\cite{Patania2023} & $\mathcal{O}((n+m)L^{3})$ \\
Cyclic graph modularity & Enumeration over $n$ cyclic cuts & $\mathcal{O}(n^{3})$~\cite{Patania2023} & $\mathcal{O}(n(n+m)L^{3})$ \\
Weighted cluster editing & $a_{i} = 1$, $\phi(x) = \eta\binom{x}{2}$, $w_{e} = \eta + b_{e}$ & $\mathcal{O}(n^{2}+m)$ & $\mathcal{O}((n+m)L^{3})$ \\
All-or-nothing HyperLam & $a_{i} = \mu_{i}$, $\phi(x) = \frac{\lambda}{2}x^{2}$, $w_{e} = 1$ & $\mathcal{O}(n^{2}+N)$ & $\mathcal{O}(N + (n+h)L^{3})$ \\
Strict hypergraph modularity & $a_{i} = d_{i}$, $\phi(x) = \sum_{\rho} h_{\rho}(x/N)^{\rho}$, $w_{e}=1$ & $\mathcal{O}(n^{2}+N)$ & $\mathcal{O}(N + (n+h)L^{3})$ \\
FENNEL ($K$ blocks) & $a_{i} = 1$, $\phi(x) = \beta x^{\gamma}$, $w_{e} = b_{e}$ & $\mathcal{O}(Kn^{2}+m)$ & $\mathcal{O}(m + K(n+m)L^{3})$ \\
Unweighted cluster editing & Endpoint degree bound $2d_{j}^{-} + 1$ & $\mathcal{O}(n^{2})$~\cite{Bellitto2015} & $\mathcal{O}(n+m)$ \\
Finite-support editing & Feasibility count $\binom{j-k}{2}$ & $\mathcal{O}(n^{2}+q)$~\cite{Bellitto2015} & $\mathcal{O}(n+q)$ \\
\bottomrule
\end{tabularx}
\end{table}

Table~\ref{tab:partition-apps} gives the running times for the following objectives, whose block costs are derived in Appendix~\ref{app:apps}.
\begin{itemize}
    \item \textbf{Ordered and cyclic graph modularity.} For a loopless undirected graph with nonnegative edge weights and total weight $W>0$, scaling modularity by $-4W^2$ gives $C(k,j)=(S_j-S_k)^2-4WE((k,j])$ with vertex masses $a_i=d_i$ and rewards $w_e=4Wb_e$~\cite{Newman2004,Patania2023}. If $b_{uv}$ is the total weight of edges between $u,v$ (zero for a missing pair), every legal cell satisfies
    \[
    \Delta C_{p,q}=2d_{p+1}d_{q+1}-4Wb_{p+1,q+1}.
    \]
Thus $s(C)\le m$, and the $m$ edge records supply the events for a sweep taking $\mathcal{O}((n+m)L^3)$ time in a linear order or $\mathcal{O}(n(n+m)L^3)$ time over the $n$ possible starting cuts of a cyclic order (Algorithm~\ref{alg:cyclic-mod}).
    \item \textbf{Weighted cluster editing.} With uniform insertion price $\eta>0$ for missing pairs and deletion price $b_e\ge0$ for edge $e$, the cost is $\sum_e b_e+\sum_{I\in\mathcal{P}}[\eta\binom{|I|}{2}-\sum_{e\subseteq I}(\eta+b_e)]$, so the algorithm runs in $\mathcal{O}((n+m)L^3)$ time for all ratios $b_e/\eta$, including the unweighted LambdaCC objective~\cite{Veldt2018} with $\eta=\lambda$ and $b_e=1-\lambda$ for $0<\lambda\le1$.
    \item \textbf{Hypergraph clustering.} For all-or-nothing HyperLam~\cite{Veldt2020} with vertex weights $\mu_i\ge0$ and resolution $\lambda\ge0$, the cost $C(k,j)=\frac{\lambda}{2}(S_j-S_k)^2-\#\{e\subseteq(k,j]\}$ gives $\mathcal{O}(N+(n+h)L^3)$ time for arbitrary hyperedge ranks. For strict hypergraph modularity~\cite{Kaminski2019} with $h_\rho$ edges of rank $\rho$, the penalty is $\phi(x)=\sum_{\rho\ge1}h_\rho(x/N)^\rho$, which reduces to $h(x/N)^\rho$ on a $\rho$-uniform hypergraph. Horner evaluation takes constant time for bounded maximum rank, giving the same $\mathcal{O}(N+(n+h)L^3)$ bound, and for unrestricted rank the running time is $\mathcal{O}(N+(n+h)L^2(L+T_\phi))$ with $T_\phi$ denoting the exact evaluation time of the polynomial.
    \item \textbf{FENNEL with $K$ blocks.} For $\beta\ge0$ and $\gamma\ge1$, minimizing $\mathrm{cutweight}(\mathcal{P})+\beta\sum_{I\in\mathcal{P}}|I|^\gamma$ over $K$ contiguous blocks takes $\mathcal{O}(m+K(n+m)L^3)$ time given a constant-time exact oracle for $\phi(x)=\beta x^\gamma$~\cite{Tsourakakis2014}, and this is a word-RAM bound for fixed integer $\gamma$ when a fixed rational scaling gives polynomially bounded integer data.
    \item \textbf{Unweighted and finite-support cluster editing.} With unit prices $b_e=\eta=1$, every optimal block ending at $j$ has at most $2d_j^-+1$ vertices, where $d_j^-$ counts the neighbors preceding $j$, giving an $\mathcal{O}(n+m)$-time algorithm (Theorem~\ref{thm:linear-unweighted-ce}). When blocks may contain only listed pairs, the number of such pairs incident to each right endpoint gives an $\mathcal{O}(n+q)$-time algorithm for a list of $q$ pairs (Theorem~\ref{thm:linear-finite-ce}).
\end{itemize}

\section{Product lower bounds and the graph of negative differences}
\label{sec:lower-bounds}

The reductions in this section bound the exponents for matrix products, vector queries, and weighted triangles in terms of the number of negative differences or the graph formed by their row and column indices.

\subsection{Reduction to a squared-distance matrix with discounts}

Partitioning the middle indices expresses a general integer min-plus product as the entrywise minimum of group products, each of which can be recovered from a product whose right factor has a squared-distance background and interval discounts.

\begin{theorem}[Reduction to a quadratic background]
\label{thm:fixed-quad-lb}
Fix $0<\alpha\le1$ and suppose a deterministic $\mathcal{O}(n^\beta)$-time algorithm computes $A\otimes B$ for $n\times n$ polynomially bounded integer matrices with
\begin{equation}
\label{eq:fixed-quad-promise}
B_{kj} = (j - k)^{2} - \sum_{\ell=1}^{r} w_{\ell}\mathbf{1}\{k \in I_{\ell}\}\mathbf{1}\{j \in J_{\ell}\}, \qquad w_{\ell} > 0, \quad r \le n^{1+\alpha},
\end{equation}
where each $I_\ell$ is a prefix or suffix, each $J_\ell$ is a suffix, $s(B)\le n^{1+\alpha}$, and all mixed differences are nonzero with $\Theta(n^2)$ of them positive. With $B$ supplied either explicitly or by the rectangle list, this algorithm gives a deterministic $\mathcal{O}(n^{1-\alpha+\beta}+n^{3-\alpha})$-time algorithm for general integer min-plus multiplication and hence for APSP with polynomially bounded integer weights.
\end{theorem}

\begin{proof}[Proof sketch, with details in Appendix~\ref{app:lower-bounds}]
For arbitrary integer matrices $A,B\in[-U,U]^{n\times n}$ and fixed $\alpha$, the finitely many sizes with $n^\alpha<4$ can be solved directly, and for larger $n$ the middle indices are partitioned into $Q=\lceil n/b\rceil=\mathcal{O}(n^{1-\alpha})$ groups of size $a\le b=\lfloor n^\alpha/4\rfloor$. For a group $\kappa_0,\ldots,\kappa_{a-1}$, place its columns of $A$ and rows of $B$ in positions $k<a$ of $A^{(g)},B^{(g)}$ and set $A^{(g)}_{ik}=2U+1$, $B^{(g)}_{kj}=0$ for $k\ge a$, so each padded path has weight $2U+1>2U$ and $(A\otimes B)_{ij}=\min_{0\le g<Q}(A^{(g)}\otimes B^{(g)})_{ij}$.

All nonzero differences $D_{pq}=\Delta B^{(g)}_{pq}$ occur in the first $a$ row gaps, so the telescoping representation $B^{(g)}_{kj}=u_k+v_j+R_{kj}$ has at most $a(n-1)\le b(n-1)$ nonzero terms in $R_{kj}=-\sum_{p<k,\,q<j}D_{pq}$. With $\lambda=2n^2+1$, the two factors are
\[
\hat{A}_{ik} = \lambda(A^{(g)}_{ik} + u_{k}) - k^{2}, \qquad B^{*}_{kj} = (j - k)^{2} + \lambda R_{kj} - F[j],
\]
where $F[j]$ collects the column terms from the identity $c\mathbf{1}\{k>p\}\mathbf{1}\{j>q\}=c\mathbf{1}\{j>q\}-c\mathbf{1}\{k\le p\}\mathbf{1}\{j>q\}$ for each positive coefficient $c$. Substitution gives
\[
\hat{A}_{ik} + B^{*}_{kj} = \lambda\bigl(A^{(g)}_{ik} + B^{(g)}_{kj}\bigr) - 2kj + \bigl(j^{2} - \lambda v_{j} - F[j]\bigr).
\]
Because $0 \le 2kj \le 2(n-1)^{2} < \lambda$ for all $k, j \in [0, n-1]$, subtracting the column potential $j^{2} - \lambda v_{j} - F[j]$ from $S_{ij} = (\hat{A} \otimes B^{*})_{ij}$ to get $\widetilde{S}_{ij}$ recovers the exact integer product entry via
\[
(A^{(g)} \otimes B^{(g)})_{ij} = \left\lceil \frac{\widetilde{S}_{ij}}{\lambda} \right\rceil.
\]
The differences $\Delta B^*_{pq}=2+\lambda D_{pq}$ are all nonzero, have negative sign precisely at the negative cells of $D$, and are positive on at least $(n-1-a)(n-1)=\Theta(n^2)$ cells, with $s(B^*)\le r\le b(n-1)\le n^{1+\alpha}$. Preparing and solving the $Q=\mathcal{O}(n^{1-\alpha})$ instances therefore takes $\mathcal{O}(n^{1-\alpha+\beta}+n^{3-\alpha})$ time.
\end{proof}

\begin{corollary}[Product exponents under $\gamma$-APSP]
\label{cor:param-lb}
For every fixed $\alpha \in (3 - c_{\mathrm{APSP}}, 1]$, any deterministic algorithm for the promise problem in Theorem~\ref{thm:fixed-quad-lb} (and hence for $A \otimes B$ under $s(B) \le n^{1+\alpha}$) has complexity exponent $\beta \ge c_{\mathrm{APSP}} - 1 + \alpha$. In particular, under the $\gamma$-APSP Hypothesis (Definition~\ref{def:gamma-apsp}), no $\mathcal{O}(n^{\gamma - 1 + \alpha - \epsilon})$-time algorithm exists for any fixed $\alpha \in (3 - \gamma, 1]$ and $\epsilon > 0$.
\end{corollary}

\begin{proof}
By Theorem~\ref{thm:fixed-quad-lb} and the definition of $c_{\mathrm{APSP}}$, we have $\max\{1-\alpha+\beta,3-\alpha\}\ge c_{\mathrm{APSP}}$, and the assumption $\alpha>3-c_{\mathrm{APSP}}$ gives $3-\alpha<c_{\mathrm{APSP}}$, so $1-\alpha+\beta\ge c_{\mathrm{APSP}}$ as claimed.
\end{proof}

For polynomially bounded integer inputs the upper exponent is at most $\min\{2+\alpha,c_{\mathrm{APSP}}\}$ by taking the faster of the sweep and a general product algorithm, and the $\Omega(n^2)$ output time together with Corollary~\ref{cor:param-lb} gives the lower exponent $\max\{2,c_{\mathrm{APSP}}-1+\alpha\}$ for all $0\le\alpha\le1$. At $\alpha=1$ the reduction uses a constant number of instances and quadratic preparation, so the class with the fixed quadratic background has exponent $c_{\mathrm{APSP}}$ in both input representations, since four corner updates per rectangle followed by a two-dimensional prefix sum materialize $B$ in $\mathcal{O}(n^2+r)$ time.

\begin{remark}[Distances in a three-layer graph with nonnegative weights]
For complete bipartite arcs between consecutive layers, the product entries are exactly the first-to-third distances, and Theorem~\ref{thm:explicit-product} returns a minimizing intermediate vertex for each pair. Shifting $A$ and $B$ uniformly by $c_{A} = \max\{0,-\min_{i,k} A_{ik}\}$ and $c_{B} = \max\{0,-\min_{k,j} B_{kj}\}$ makes all edge weights nonnegative and adds $c_{A}+c_{B}$ to every first-to-third distance. The shift preserves $\Delta B$, so the exponent lower bound of Corollary~\ref{cor:param-lb} also holds for computing these distances under the same negative-support bound. For the instances in~\eqref{eq:fixed-quad-promise}, the shifted quadratic background is $(j-k)^2+c_B$.
\end{remark}

\subsection{The negative-support graph and repeated products}

The bipartite \emph{negative-support graph} $G_{-}(B)$ has $n-1$ row gaps and $n-1$ column gaps as vertices, with an edge $(p,q)$ whenever $\Delta B_{p,q} < 0$. Let $s = |E(G_{-}(B))|$, $\tau = \mathrm{vc}(G_{-}(B))$ (vertex cover number), $\Delta_{-} = \max_{v} \deg_{G_{-}(B)}(v)$ (maximum degree), and $t = \mathrm{tw}(G_{-}(B))$ (treewidth). Since $s \le (n-1)\Delta_{-}$ and a graph of treewidth $t$ on $2(n-1)$ vertices is $t$-degenerate ($s \le 2(n-1)t$), Theorem~\ref{thm:explicit-product} gives $\tilde{\mathcal{O}}(n^{2}(\Delta_{-} + 1))$ and $\tilde{\mathcal{O}}(n^{2}(t + 1))$ multiplication bounds. Cutting at the gaps of a vertex cover of size at most $2\tau$ partitions $B$ into Monge submatrices on which SMAWK computes column minima.

Global row and column cuts partition $B$ into Monge rectangles if and only if the chosen gaps cover every edge of $G_-(B)$, since an uncovered negative cell lies inside a block and violates its adjacent inequality, and telescoping the nonnegative differences of a block gives the Monge inequality throughout that block. The minimum number of such cuts is therefore $\tau$.

For $n\ge2$ and a permutation $\sigma$ of $[0,n-2]$, consider the matrix
\[
 P(i,j)=|\{p<i:\sigma(p)<j\}|,\qquad B_{ij}=-ij+2P(i,j).
\]
Here $\Delta B_{p,q}=1-2\mathbf{1}\{q=\sigma(p)\}$, so all $(n-1)^2$ differences are nonzero and the negative-support graph is a perfect matching with $s=\tau=n-1$ and $\Delta_-=t=1$. More generally, the matrix $B_{ij}=-\sum_{p<i,\,q<j}F_{p,q}$ realizes any specified mixed-difference array $F$.

\begin{theorem}[Product bounds from the negative-support graph]
\label{thm:support-graph}
\leavevmode
\begin{enumerate}
    \item \textbf{Vertex cover.} Given explicit $n\times n$ matrices $A,B$, Algorithm~\ref{alg:gap-cuts} uses the endpoints of a maximal matching as a vertex cover and computes $A\otimes B$ in $\mathcal{O}(n^2(\tau(B)+1))$ time and $\mathcal{O}(n^2)$ space.
    \item \textbf{Exponent lower bounds.} For every fixed $\alpha \in (3 - c_{\mathrm{APSP}}, 1]$, any deterministic algorithm for $A \otimes B$ under any single promise $s(B) \le n^{1+\alpha}$, $\tau(B) \le n^{\alpha}$, $\Delta_{-}(B) \le n^{\alpha}$, or $t(B) \le n^{\alpha}$ requires exponent $\beta \ge c_{\mathrm{APSP}} - 1 + \alpha$.
\end{enumerate}
\end{theorem}

The algorithms for support size, maximum degree, and treewidth obtain their input from a scan of the mixed differences, and in each case the running time can be bounded by the smaller of the stated bound and the time for direct cubic multiplication.

For arbitrary finite $X_0$, Algorithm~\ref{alg:repeated-mult} computes $X_0\otimes B_1\otimes\cdots\otimes B_\ell$ from left to right in $\mathcal{O}(\ell n^2L+nL^3\sum_{t=1}^{\ell}s(B_t))$ time and $\mathcal{O}(n^2)$ working space beyond the supplied matrices. Storing the witnesses at each layer takes another $\mathcal{O}(\ell n^2)$ space and permits path reconstruction for every source and terminal, including when an intermediate product has the quadratic negative support in the following example.

\begin{example}[Negative differences in a product]
\label{ex:support-blowup}
For $0 \le i, k, j < n$, define
\[
A_{ik} = \begin{cases} i & k = 0, \\ 0 & k > 0, \end{cases}
\qquad
B_{kj} = \begin{cases} n & k > 0 \text{ and } j \text{ odd}, \\ 0 & \text{otherwise.} \end{cases}
\]
The matrix $A$ is Monge and $B$ has $s(B)=\lfloor n/2\rfloor$ negative differences on its first row gap, while the product $(A\otimes B)_{ij}=i\mathbf{1}\{j\text{ odd}\}$ has differences alternating between $-1$ and $+1$ along each row gap, giving $s(A\otimes B)=(n-1)\lfloor n/2\rfloor=\Theta(n^2)$.
\end{example}

\subsection{Vector queries and weighted triangles}

\begin{corollary}[Amortized vector queries]
\label{cor:adaptive-query-lower}
For fixed $3-c_{\mathrm{APSP}}<\alpha\le1$, consider a deterministic data structure that preprocesses a polynomially bounded integer matrix $B$ with $s(B)\le n^{1+\alpha}$ in $\mathcal{O}(n^2)$ time and answers $n$ products with polynomially bounded integer vectors in amortized time $\mathcal{O}(n^\eta)$ per vector. Its query exponent satisfies $\eta\ge c_{\mathrm{APSP}}-2+\alpha$, including for the right factors of Theorem~\ref{thm:fixed-quad-lb}.
\end{corollary}
\begin{proof}
Querying the $n$ rows of an arbitrary polynomially bounded integer matrix $A$ computes $A\otimes B$ in $\mathcal{O}(n^2+n^{1+\eta})$ time, so Corollary~\ref{cor:param-lb} and $c_{\mathrm{APSP}}-1+\alpha>2$ give $1+\eta\ge c_{\mathrm{APSP}}-1+\alpha$. These queries are fixed in advance and hence form a valid query sequence for a structure that also accepts adaptive queries.
\end{proof}

For explicit matrices $A,B,Z$, write
\[
 \operatorname{Tri}(A,B,Z)=\min_{i,k,j}(A_{ik}+B_{kj}+Z_{ji})
                         =\min_{i,j}\bigl((A\otimes B)_{ij}+Z_{ji}\bigr).
\]
Minimum-Weight Triangle returns this value and Negative Triangle tests its sign, so scanning the $n^2$ entries after multiplication solves either problem in $\mathcal{O}(n^2L+nsL^3)$ or $\mathcal{O}(n^2(\tau+1))$ time and $\mathcal{O}(n^2)$ space. The case $A=0$, $B_{kj}=(j-k)^2$ has $s=\tau=\Delta_-=t=0$ and every product entry equal to zero, so on an all-zero $Z$ a deterministic algorithm must inspect all $n^2$ entries to distinguish this input from one with a single negative entry.

Let $c_{\triangle}$ be the infimum of the deterministic time exponents for general Negative Triangle with polynomially bounded integer weights, so computing a full product and scanning it gives $c_{\triangle}\le c_{\mathrm{APSP}}$. Vassilevska Williams and Williams~\cite{Williams2018} relate APSP and Negative Triangle by reductions preserving truly subcubic time, and the two triangle problems have the same exponent because subtracting a threshold from $Z$ tests whether the optimum is below that threshold, allowing binary search over the integer weight range with $\mathcal{O}(\log n)$ sign tests.

\begin{proposition}[Exponent transfer for weighted triangles]
\label{prop:triangle-transfer}
Fix $0<\alpha\le1$ and impose any one of the four promises on $B$ in Theorem~\ref{thm:support-graph}. A deterministic $\mathcal{O}(n^\beta)$ algorithm for either triangle problem under that promise gives a deterministic $\mathcal{O}(n^{1-\alpha+\beta}+n^{3-\alpha})$ algorithm for general Negative Triangle. A restricted minimum-weight solver also computes the general minimum weight within the same time bound. For the promise $s(B)\le n^{1+\alpha}$ this also holds with the fixed quadratic background and interval discounts of Theorem~\ref{thm:fixed-quad-lb}, in either input representation. Consequently $\beta\ge c_{\triangle}-1+\alpha$ whenever $3-c_{\triangle}<\alpha\le1$.
\end{proposition}
\begin{proof}
For a bound $U$ on the absolute entries of $A,B,Z$, use the groups of Theorem~\ref{thm:fixed-quad-lb} with padding $H=5U+1$ in $A^{(g)}$ and zero in $B^{(g)}$, so a padded triangle weighs at least $H-U>3U$ and the minimum over the groups equals the original optimum. For the fixed quadratic background use the factors $\hat A,B^*$ and quantities $\lambda,v_j,F[j]$ from the product reduction, with third factor
\[
 Z^*_{ji}=\lambda(Z_{ji}+v_j)-j^2+F[j]+2n^2.
\]
A triangle of integer weight $T=A^{(g)}_{ik}+B^{(g)}_{kj}+Z_{ji}$ now weighs $\lambda T+2n^2-2kj$, where $0\le2n^2-2kj<\lambda$, so the transformed sum is negative exactly for $T<0$, is nonnegative for $T=0$, and gives the group optimum $\lfloor\operatorname{Tri}(\hat A,B^*,Z^*)/\lambda\rfloor$ after minimization. The first $a\le\lfloor n^\alpha/4\rfloor$ row gaps cover all negative cells, and taking those gaps together with one successive column gap in each bag gives a path decomposition of width at most $a$, with singleton bags for isolated vertices.

For the maximum-degree promise, the cyclic block construction in the proof of Theorem~\ref{thm:support-graph} with filler $H=5U+1$ retains each original entry of $B$ in one block and gives every triangle using a filler weight at least $H-2U>3U$. Each construction prepares $\mathcal{O}(n^{1-\alpha})$ instances in $\mathcal{O}(n^2)$ time per instance, and the disjunction of the sign tests gives the decision answer while the least recovered group optimum gives the exact answer. When $\alpha>3-c_{\triangle}$ the preparation exponent $3-\alpha$ is below $c_{\triangle}$, giving $1-\alpha+\beta\ge c_{\triangle}$ from the definition of the exponent.
\end{proof}

For bounded-error algorithms the same transfer holds with the randomized Negative Triangle exponent, since $\mathcal{O}(\log n)$ independent repetitions per instance bound the error probability of the disjunction by $1/3$ and preserve the time exponents.

\section{Conclusions and open problems}

For an arbitrary left factor, $s(B)=\mathcal{O}(n)$ gives a near-quadratic min-plus product, and a constant-time Monge oracle with $\mathcal{O}(n)$ rectangle corrections gives near-linear sequential partitioning, including the convex-load objectives on sparse ordered graphs. For $s(B)\le n^{1+\alpha}$ with $3-c_{\mathrm{APSP}}<\alpha\le1$, the sweep has upper exponent $2+\alpha$ and the squared-distance reduction gives lower exponent $c_{\mathrm{APSP}}-1+\alpha$, with equality to the general product exponent at $\alpha=1$ as shown in Section~\ref{sec:lower-bounds}.

The running time of a single recurrence and the cost of successive sweeps give the following open problems. Appendix~\ref{app:open} discusses the input representation and the choice of order.
\begin{enumerate}
    \item \textbf{Dependence on the number of rectangles.} Classical Monge recurrences take $\mathcal{O}(n)$ time when $r=0$~\cite{Wilber1988, Larmore1991}, and an $\mathcal{O}(n+r\operatorname{polylog}n)$ bound for rectangle corrections would retain this linear term, including for internal rewards whose events at $j=\max e$ affect rows $k<\min e$.
    \item \textbf{Lower bounds for one recurrence.} Theorem~\ref{thm:fixed-quad-lb} bounds the exponent for a batch of matrix products, and the corresponding problem for a single recurrence~\eqref{eq:dp-unrestricted} is to determine a conditional lower bound in terms of $n$ and the number $r$ of supplied rectangles.
    \item \textbf{Vertex cover and treewidth in sequential DP.} A bound in terms of the vertex cover number $\tau(C)$ or treewidth $t(C)$ of the negative-support graph would extend Theorem~\ref{thm:support-graph} to a triangular recurrence, where the order of the block computations must respect the availability of the prefix values $D[k]$.
    \item \textbf{Cyclic orders and successive layers.} Each rotation of a cyclic order in Algorithm~\ref{alg:cyclic-mod} and each layer of fixed-$K$ partitioning in Algorithm~\ref{alg:k-blocks} starts a new sweep, leaving open whether the envelope information from one run can reduce the total work over all rotations or layers.
\end{enumerate}

\appendix

\section{Pseudocode for the column sweep}
\label{app:data-structure}

In the balanced tree $T$ of Section~\ref{sec:engine}, each node $u$ stores its interval $K(u)$ and the number of active leaves, and each internal node stores a lazy addition $z_u$ and switch column $\theta_u$. For a nonempty node, $\operatorname{APPLY}(u,\beta)$ adds $\beta$ to the stored potential $v_k$ at a leaf or to the tag $z_u$ at an internal node, and for an empty node the operation returns immediately. The operation $\operatorname{PUSH}(u)$ applies $z_u$ to both children and then sets $z_u$ to zero.

\begin{algorithmsteps}{eval-pull}{Envelope evaluation and switch columns}
1: & \textbf{function} $\operatorname{EVAL}(u, x)$ \\*
2: & \quad \textbf{if} $u$ is empty \textbf{then return} $(+\infty, -1)$ \textbf{end if} \\
3: & \quad \textbf{if} $u$ is the leaf for $k$ \textbf{then return} $(v_{k} + M(k,x), k)$ \textbf{end if} \\
4: & \quad \textbf{if} $x < \theta_{u}$ \textbf{then} $(val, idx) \leftarrow \operatorname{EVAL}(\mathrm{left}(u), x)$ \textbf{else} $(val, idx) \leftarrow \operatorname{EVAL}(\mathrm{right}(u), x)$ \textbf{end if} \\
5: & \quad \textbf{return} $(val + z_{u}, idx)$ \\
6: & \textbf{end function} \\
\addlinespace
7: & \textbf{function} $\operatorname{PULL}(u, j)$ \\
8: & \quad $a \leftarrow \mathrm{left}(u)$; $b \leftarrow \mathrm{right}(u)$; recompute active count of $u$ from $a, b$ \\
9: & \quad \textbf{if} $b$ is empty \textbf{then} $\theta_{u} \leftarrow \nu + 1$; \textbf{return} \textbf{end if} \\
10: & \quad \textbf{if} $a$ is empty \textbf{then} $\theta_{u} \leftarrow j$; \textbf{return} \textbf{end if} \\
11: & \quad $l \leftarrow j$; $h \leftarrow \nu + 1$ \\
12: & \quad \textbf{while} $l < h$ \textbf{do} \\
13: & \quad \quad $x \leftarrow \lfloor (l+h)/2 \rfloor$ \\
14: & \quad \quad \textbf{if} $\operatorname{EVAL}(b,x).\mathrm{value} < \operatorname{EVAL}(a,x).\mathrm{value}$ \textbf{then} $h \leftarrow x$ \textbf{else} $l \leftarrow x + 1$ \textbf{end if} \\
15: & \quad \textbf{end while} \\
16: & \quad $\theta_{u} \leftarrow l$ \\*
17: & \textbf{end function} \\
\end{algorithmsteps}

\begin{algorithmsteps}{add-activate}{Interval addition and row activation}
1: & \textbf{function} $\operatorname{ADD}(u, I, \beta, j)$ \\*
2: & \quad \textbf{if} $K(u) \cap I = \emptyset$ or $u$ is empty \textbf{then return} \textbf{end if} \\
3: & \quad \textbf{if} $K(u) \subseteq I$ \textbf{then} $\operatorname{APPLY}(u, \beta)$; \textbf{return} \textbf{end if} \\
4: & \quad $\operatorname{PUSH}(u)$; $\operatorname{ADD}(\mathrm{left}(u), I, \beta, j)$; $\operatorname{ADD}(\mathrm{right}(u), I, \beta, j)$; $\operatorname{PULL}(u, j)$ \\
5: & \textbf{end function} \\
\addlinespace
6: & \textbf{function} $\operatorname{ACTIVATE}(u, k, v, j)$ \\
7: & \quad \textbf{if} $u$ is a leaf \textbf{then} $v_{k} \leftarrow v$; set active count to $1$; \textbf{return} \textbf{end if} \\
8: & \quad $\operatorname{PUSH}(u)$; let $c$ be the child of $u$ containing $k$ \\
9: & \quad $\operatorname{ACTIVATE}(c, k, v, j)$; $\operatorname{PULL}(u, j)$ \\*
10: & \textbf{end function} \\
\end{algorithmsteps}

\begin{algorithmsteps}{static-transfer}{Vector product $\operatorname{STATICTRANSFER}(x,M,E)$}
1: & $\nu \leftarrow n - 1$; build balanced binary tree $T$ on $[0, n-1]$ \\*
2: & At leaf $k$, set $v_{k} \leftarrow x_{k}$ and mark active (if $x_{k} = +\infty$, leave inactive); set all $z_{u} \leftarrow 0$ \\
3: & \textbf{for} each internal node $u$ in bottom-up order \textbf{do} $\operatorname{PULL}(u, 0)$ \textbf{end for} \\
4: & \textbf{for} $j = 0, \dots, n-1$ \textbf{do} \\
5: & \quad \textbf{for} each event $(I, \beta)$ in bucket $E[j]$ \textbf{do} $\operatorname{ADD}(T.\mathrm{root}, I, \beta, j)$ \textbf{end for} \\
6: & \quad $(y_{j}, \pi_{j}) \leftarrow \operatorname{EVAL}(T.\mathrm{root}, j)$ \\
7: & \textbf{end for} \\*
8: & \textbf{return} $y, \pi$ \\
\end{algorithmsteps}

\begin{algorithmsteps}{seq-dp}{Sequential partitioning with rectangle corrections}
1: & $\nu \leftarrow n$; bucket entry $(I_{\ell}, +\alpha_{\ell})$ at $a_{\ell}$ and exit $(I_{\ell}, -\alpha_{\ell})$ at $b_{\ell}+1$ (if $b_{\ell} < n$) \\*
2: & Initialize empty envelope tree $T$ and zero difference-array Fenwick tree $H$ on $[0, n-1]$ \\
3: & $D[0] \leftarrow 0$ \\
4: & \textbf{for} $j = 1, \dots, n$ \textbf{do} \\
5: & \quad \textbf{for} each event $(I, \beta)$ in bucket $j$ \textbf{do} \\
6: & \quad \quad $H.\mathrm{add}(I, \beta)$; $\operatorname{ADD}(T.\mathrm{root}, I \cap [0, j-2], \beta, j)$ \\
7: & \quad \textbf{end for} \\
8: & \quad $k \leftarrow j - 1$ \\
9: & \quad \textbf{if} $D[k] < +\infty$ \textbf{then} $\operatorname{ACTIVATE}(T.\mathrm{root}, k, D[k] + H.\mathrm{point}(k), j)$ \textbf{end if} \\
10: & \quad $(D[j], \pi[j]) \leftarrow$ minimum at column $j$ over $F_{j}$ (default $F_{j} = [0, j-1]$) \\
11: & \textbf{end for} \\*
12: & \textbf{return} $(D, \pi)$ \\
\end{algorithmsteps}

\begin{algorithmsteps}{k-blocks}{Partition into $K$ blocks}
1: & $P[0] \leftarrow 0$; $P[1..n] \leftarrow +\infty$ \\*
2: & \textbf{for} $b = 1, \dots, K$ \textbf{do} \\
3: & \quad Initialize $Q[0..n] \leftarrow +\infty$ \\
4: & \quad Run Algorithm~\ref{alg:seq-dp}, activating row $k$ with $P[k] + H.\mathrm{point}(k)$ whenever $P[k] < +\infty$, writing answers to $Q[j]$ and predecessors to $\pi_{b}[j]$ \\
5: & \quad $P \leftarrow Q$ \\
6: & \textbf{end for} \\*
7: & \textbf{return} $P[n]$ and partition reconstructed via $(b, j) \leftarrow (b-1, \pi_{b}[j])$ \\
\end{algorithmsteps}

\begin{algorithmsteps}{persistent-oracle}{Monge oracle from a list of negative cells}
1: & Bucket negative cells $(p, q, w_{p,q})$ by column $q$; $V_{0} \leftarrow$ zero segment tree on $[0, n-1]$ \\*
2: & \textbf{for} $j = 1, \dots, n$ \textbf{do} \\
3: & \quad $V_{j} \leftarrow V_{j-1}$ \\
4: & \quad \textbf{for} each $(p, j-1, w)$ in bucket $j-1$ \textbf{do} $V_{j} \leftarrow \operatorname{PersistentPointAdd}(V_{j}, p, w)$ \textbf{end for} \\
5: & \textbf{end for} \\
6: & Define $M(k,j) = C(k,j) + \operatorname{SuffixSum}(V_{j}, k)$ and create rectangles $([0,p], [q+1, n], -w_{p,q})$ \\*
7: & \textbf{return} output of Algorithm~\ref{alg:seq-dp} with background $M$ and these rectangles \\
\end{algorithmsteps}

\section{Algorithms for contiguous partitions and convex transitions}
\label{app:apps}

\begin{algorithmsteps}{convex-chain}{Layered recurrence with convex transitions}
1: & $d \leftarrow d_{0}$ \\*
2: & \textbf{for} $t = 1, \dots, \ell$ \textbf{do} \\
3: & \quad Bucket $(I, \alpha)$ at $a$ and $(I, -\alpha)$ at $b+1$ ($b+1 < n$) for $B_{t}$ to form $E_{t}$; define $M_{t}(k,j) = \phi_{t}(y^{(t)}_{j} - x^{(t)}_{k})$ \\
4: & \quad $(z, \pi_{t}) \leftarrow \operatorname{STATICTRANSFER}(d, M_{t}, E_{t})$ (Algorithm~\ref{alg:static-transfer}) \\
5: & \quad $d(j) \leftarrow z(j) + u_{t}(j)$ for all $j \in [0, n-1]$; retain $\pi_{t}$ if reconstruction is required \\
6: & \textbf{end for} \\*
7: & \textbf{return} $d$ and retained predecessor arrays \\
\end{algorithmsteps}

\begin{algorithmsteps}{interaction-dp}{Contiguous partitions with internal rewards}
1: & Compute prefix sums $S_{j} = \sum_{i=1}^{j} a_{i}$ ($S_{0} = 0$) \\*
2: & \textbf{for} each interaction $e \in \mathcal{E}$ \textbf{do} compute $u_{e} = \min e$, $v_{e} = \max e$, and append rectangle $([0, u_{e}-1], [v_{e}, n], -w_{e})$ \textbf{end for} \\*
3: & Define $M(k,j) = \phi(S_{j} - S_{k})$ and run Algorithm~\ref{alg:seq-dp} (or Algorithm~\ref{alg:k-blocks} for $K$ blocks) \\
\end{algorithmsteps}

\paragraph{Graph modularity.} For a loopless undirected graph $G=([1,n],E)$ with edge weights $b_e\ge0$, let $W=\sum_e b_e>0$ and $d_i=\sum_{e\ni i}b_e$. Multiplying $Q(\mathcal{P})=\sum_{I\in\mathcal{P}}(E(I)/W-\mathrm{vol}(I)^2/(4W^2))$ by $-4W^2$ gives $C(k,j)=(S_j-S_k)^2-4WE((k,j])$, so Theorem~\ref{thm:interaction-dp} with $a_i=d_i$, $\phi(x)=x^2$ and $w_e=4Wb_e$ computes an optimal linear partition in $\mathcal{O}((n+m)L^3)$ time.

\besidetext{.45\linewidth}{%
    \noindent\algheading{cyclic-mod}\enspace Cyclic graph modularity\par
    \smallskip
    \begin{tabularx}{\linewidth}{@{}r@{\enspace}>{\raggedright\arraybackslash}X@{}}
        \toprule
        1. & Set $Q^*\gets-\infty$. \\
        2. & For each gap, rotate the order so that vertex $1$ follows the gap. \\
        3. & For this rotation, run Algorithm~\ref{alg:interaction-dp} with $a_i=d_i$, $\phi(x)=x^2$, $w_e=4Wb_e$. \\
        4. & Retain the best $-D[n]/(4W^2)$ and its partition in the cyclic order. \\
        5. & Return the retained partition. \\
        \bottomrule
    \end{tabularx}\par
}{%
    In a cyclic order, every partition with at least two blocks has a boundary at one of the $n$ gaps, and a partition consisting of a single block is feasible for every starting cut. Choosing a boundary as the start of the linear order puts each block of the cyclic partition in a single interval, so the best of the $n$ linear optima is the cyclic optimum. Each rotation returns its modularity as $-D[n]/(4W^2)$, where $D[n]$ is the minimum block cost for the rotated order.
}
Algorithm~\ref{alg:cyclic-mod} translates the selected cuts back to the given order and reuses the storage on each rotation, giving $\mathcal{O}(n(n+m)L^3)$ time and $\mathcal{O}(n+m)$ space.

\paragraph{Weighted cluster editing.} Relative to the cost $\sum_e b_e$ of deleting every edge, a block $I$ pays $\eta$ for each missing pair and saves $b_e$ on each retained edge, giving a contribution $\eta\binom{|I|}{2}-\sum_{e\subseteq I}(\eta+b_e)$, so Theorem~\ref{thm:interaction-dp} applies with $a_i=1$, $\phi(x)=\eta x(x-1)/2$ and $w_e=\eta+b_e$ in $\mathcal{O}((n+m)L^3)$ time.

\paragraph{Hypergraph objectives.} For all-or-nothing HyperLam~\cite{Veldt2020}, the objective $\mathrm{cutnets}(\mathcal{P})+\lambda\sum_{u<v\text{ co-clustered}}\mu_u\mu_v$ equals $h-\frac{\lambda}{2}\sum_i\mu_i^2+\sum_{I\in\mathcal{P}}[\frac{\lambda}{2}(\sum_{i\in I}\mu_i)^2-\#\{e\subseteq I\}]$, so Theorem~\ref{thm:interaction-dp} with $a_i=\mu_i$, $\phi(x)=\lambda x^2/2$ and $w_e=1$ gives $\mathcal{O}(N+(n+h)L^3)$ time for arbitrary ranks. For strict hypergraph modularity~\cite{Kaminski2019}, the incidence degrees $a_i=d_i$ have sum $N$, and the objective is
\[
 Q_H(\mathcal P)=\frac{1}{h}\sum_{I\in\mathcal P}
 \left[\#\{e\subseteq I\}-\sum_{\rho\ge1} h_\rho\bigl(\mathrm{vol}(I)/N\bigr)^\rho\right].
\]
The polynomial $\phi(x)=\sum_{\rho\ge1}h_\rho(x/N)^\rho$ is convex on $\mathbb{R}_{\ge0}$ because its coefficients are nonnegative, and when $\rho_{\max}=\mathcal{O}(1)$ its constant evaluation time gives $\mathcal{O}(N+(n+h)L^3)$ total time. At bounded rank, multiplying by $N^{\rho_{\max}}$ gives integer costs of polynomial magnitude for polynomially bounded incidence size, so the computation is exact on the word RAM.

\paragraph{FENNEL with $K$ blocks.} Since $\mathrm{cutweight}(\mathcal{P})=W-\sum_{I\in\mathcal{P}}E(I)$, Theorem~\ref{thm:interaction-dp} with $a_i=1$, $\phi(x)=\beta x^\gamma$ and $w_e=b_e$ gives $\mathcal{O}(m+K(n+m)L^3)$ time for $K$ blocks when the penalty takes constant time to evaluate.

\begin{theorem}[Unweighted contiguous cluster editing]
\label{thm:linear-unweighted-ce}
For a simple graph on $n$ ordered vertices with $m$ edges, unweighted contiguous cluster editing ($\eta = b_{e} = 1$) can be solved in $\mathcal{O}(n+m)$ time and $\mathcal{O}(n)$ working space beyond the edges, bucketed by their larger endpoint (Algorithm~\ref{alg:linear-unweighted}).
\end{theorem}

\begin{proof}
Let $d_j^-$ count the neighbors of $j$ preceding it and let $d_I(j)$ count its neighbors in the last block $I$ of an optimal partition of $[1,j]$. If $|I|=\ell$, splitting off $\{j\}$ changes the cost by $2d_I(j)-(\ell-1)$, so optimality gives $\ell-1\le2d_I(j)\le2d_j^-$~\cite{Lin2015}. With edges grouped by their larger endpoint $j$, maintaining $B[u]=\#\{\text{edges }(u,v):u<v\le j\}$ and scanning $k$ down from $j-1$ to $\max\{0,j-2d_j^--1\}$ evaluates at most $\sum_{j=1}^n(2d_j^-+1)=n+2m$ candidates in $\mathcal{O}(n+m)$ total time.
\end{proof}

\begin{theorem}[Finite-support weighted contiguous cluster editing]
\label{thm:linear-finite-ce}
Let $\mathcal{L}$ consist of $q$ distinct unordered pairs $\{u,v\}$ with finite signed weights $w_{uv}$, including zero weights, and let a block be feasible when every pair of its vertices belongs to $\mathcal{L}$. With $W_+=\sum_{\{u,v\}\in\mathcal{L}}\max\{w_{uv},0\}$, Algorithm~\ref{alg:linear-finite} minimizes
\[
 W_+-\sum_{I\in\mathcal{P}}\ \sum_{\substack{\{u,v\}\in\mathcal{L}\\ u,v\in I}}w_{uv}
\]
over feasible contiguous partitions in $\mathcal{O}(n+q)$ time using $\mathcal{O}(n)$ working space beyond the listed pairs, bucketed by their larger endpoint.
\end{theorem}

\begin{proof}
A feasible block $(k,j]$ contains only listed pairs, so if $f_j^-$ counts the listed pairs joining $j$ to an earlier vertex, its size satisfies $j-k\le f_j^-+1$ and its internal pair count equals $\binom{j-k}{2}$. Scanning $k$ down from $j-1$ to $\max\{0,j-f_j^--1\}$ and testing this count examines at most $\sum_{j=1}^n(f_j^-+1)=n+q$ candidates in $\mathcal{O}(n+q)$ time.
\end{proof}

\begin{algorithmsteps}{linear-unweighted}{Unweighted contiguous cluster editing}
1: & Bucket edges $(u, j)$ ($u < j$) by $j$; $B[1..n] \leftarrow 0$; $D[0] \leftarrow 0$ \\*
2: & \textbf{for} $j = 1, \dots, n$ \textbf{do} \\
3: & \quad $d \leftarrow$ size of bucket $j$; \textbf{for} each $(u, j)$ in bucket $j$ \textbf{do} $B[u] \leftarrow B[u] + 1$ \textbf{end for} \\
4: & \quad $e \leftarrow 0$; $D[j] \leftarrow +\infty$ \\
5: & \quad \textbf{for} $k = j-1, j-2, \dots, \max\{0, j - 2d - 1\}$ \textbf{do} \\
6: & \quad \quad $e \leftarrow e + B[k+1]$; update $(D[j], \pi[j])$ with $(D[k] + \binom{j-k}{2} - 2e, k)$ \\
7: & \quad \textbf{end for} \\
8: & \textbf{end for} \\*
9: & \textbf{return} $m + D[n]$ and partition reconstructed from $\pi$ \\
\end{algorithmsteps}

\begin{algorithmsteps}{linear-finite}{Finite-support contiguous cluster editing}
1: & Bucket listed pairs $(u, j, w)$ ($u < j$) by $j$; $B_{c}[1..n] \leftarrow 0$; $B_{w}[1..n] \leftarrow 0$; $D[0] \leftarrow 0$ \\*
2: & \textbf{for} $j = 1, \dots, n$ \textbf{do} \\
3: & \quad $f \leftarrow$ size of bucket $j$; \textbf{for} each $(u, j, w)$ in bucket $j$ \textbf{do} $B_{c}[u] \leftarrow B_{c}[u] + 1$; $B_{w}[u] \leftarrow B_{w}[u] + w$ \textbf{end for} \\
4: & \quad $c \leftarrow 0$; $z \leftarrow 0$; $D[j] \leftarrow +\infty$ \\
5: & \quad \textbf{for} $k = j-1, j-2, \dots, \max\{0, j - f - 1\}$ \textbf{do} \\
6: & \quad \quad $c \leftarrow c + B_{c}[k+1]$; $z \leftarrow z + B_{w}[k+1]$ \\
7: & \quad \quad \textbf{if} $c = \binom{j-k}{2}$ \textbf{then} update $(D[j], \pi[j])$ with $(D[k] - z, k)$ \textbf{end if} \\
8: & \quad \textbf{end for} \\
9: & \textbf{end for} \\*
10: & \textbf{return} $W_{+} + D[n]$ and partition reconstructed from $\pi$ \\
\end{algorithmsteps}

\section{Matrix product algorithms and proofs}
\label{app:lower-bounds}

\begin{algorithmsteps}{explicit-product}{Matrix product from negative differences}
1: & Initialize empty buckets $E_{0}, \dots, E_{n-1}$; $R(n-1, j) \leftarrow 0$, $R(k, 0) \leftarrow 0$ \\*
2: & \textbf{for} $p, q = 0, \dots, n-2$ \textbf{do} $w_{p,q} \leftarrow \max\{-\Delta B_{p,q}, 0\}$; \textbf{if} $w_{p,q} > 0$ \textbf{then} append $([0,p], -w_{p,q})$ to $E_{q+1}$ \textbf{end if} \textbf{end for} \\
3: & \textbf{for} $k = n-2, \dots, 0$ \textbf{do for} $j = 1, \dots, n-1$ \textbf{do} $R(k,j) \leftarrow R(k+1,j) + R(k,j-1) - R(k+1,j-1) + w_{k,j-1}$ \textbf{end for end for} \\
4: & $M \leftarrow B + R$; \textbf{for} $i = 0, \dots, n-1$ \textbf{do} $(C_{i,*}, W_{i,*}) \leftarrow \operatorname{STATICTRANSFER}(A_{i,*}, M, E)$ \textbf{end for} \\*
5: & \textbf{return} $C, W$ \\
\end{algorithmsteps}

\begin{algorithmsteps}{fixed-quad-red}{Reduction to a squared-distance matrix with discounts}
1: & $U \leftarrow \max$ absolute entry of $A, B$; $H \leftarrow 2U + 1$; $b \leftarrow \lfloor n^{\alpha}/4 \rfloor$; $\lambda \leftarrow 2n^{2} + 1$; initialize $C_{ij} \leftarrow +\infty$ \\*
2: & If $b=0$, return the direct product. Otherwise partition $[0,n-1]$ into groups of size $a\le b$. \\
3: & \textbf{for} each group $\kappa_{0}, \dots, \kappa_{a-1}$ \textbf{do} \\
4: & \quad Define $A^{(g)}_{ik} = A_{i,\kappa_{k}}$ ($k < a$) else $H$; $B^{(g)}_{kj} = B_{\kappa_{k}, j}$ ($k < a$) else $0$ \\
5: & \quad $u_{k} \leftarrow B^{(g)}_{k,0}$; $v_{j} \leftarrow B^{(g)}_{0,j} - B^{(g)}_{0,0}$; $D_{pq} \leftarrow \Delta B^{(g)}_{pq}$; $\hat{A}_{ik} \leftarrow \lambda(A^{(g)}_{ik} + u_{k}) - k^{2}$ \\
6: & \quad Initialize discount list $\mathcal{R} \leftarrow \emptyset$ and difference array $F[0..n] \leftarrow 0$ \\
7: & \quad \textbf{for} each $(p,q)$ with $D_{pq} \neq 0$ \textbf{do} \\
8: & \quad \quad $c \leftarrow -\lambda D_{pq}$ \\
9: & \quad \quad \textbf{if} $c < 0$ \textbf{then} append $([p+1, n-1], [q+1, n-1], -c)$ to $\mathcal{R}$ \\
10: & \quad \quad \textbf{else} append $([0, p], [q+1, n-1], c)$ to $\mathcal{R}$; $F[q+1] \leftarrow F[q+1] + c$ \textbf{end if} \\
11: & \quad \textbf{end for} \\
12: & \quad Prefix-sum $F$. Use $B^{*}_{kj} = (j-k)^{2} - \sum_{(I,J,w)\in\mathcal{R}} w\mathbf{1}\{k \in I\}\mathbf{1}\{j \in J\}$, materializing it if needed. \\
13: & \quad Query solver for $S \leftarrow \hat{A} \otimes B^{*}$; for all $i, j$, set $C_{ij} \leftarrow \min\{C_{ij}, \lceil (S_{ij} - j^{2} + \lambda v_{j} + F[j])/\lambda \rceil\}$ \\
14: & \textbf{end for} \\*
15: & \textbf{return} $C$ \\
\end{algorithmsteps}

\begin{proof}[Proof of Theorem~\ref{thm:fixed-quad-lb}]
For inputs $A,B$ with entries in $[-U,U]$, double telescoping in a group of size $a\le b=\lfloor n^\alpha/4\rfloor$ of Algorithm~\ref{alg:fixed-quad-red} gives $B^{(g)}_{kj}=u_k+v_j+R_{kj}$ with $R_{kj}=-\sum_{p<k,\,q<j}D_{pq}$. Since $B^{(g)}_{kj}=0$ for $k\ge a$, every nonzero $D_{pq}$ has $p<a$ and there are at most $a(n-1)\le b(n-1)$ such terms, whose two sign cases give
\[
\lambda R_{kj} = F[j] - \sum_{(I,J,w)\in\mathcal{R}} w\mathbf{1}\{k \in I\}\mathbf{1}\{j \in J\}.
\]
The listed discounts therefore give $B^*_{kj}=(j-k)^2+\lambda R_{kj}-F[j]$, and the identity $R_{kj}=B^{(g)}_{kj}-u_k-v_j$ evaluates each explicit entry in constant time. For every $i,k,j\in[0,n-1]$ the sum of the two factor entries is
\[
\hat{A}_{ik} + B^{*}_{kj} = \lambda\bigl(A^{(g)}_{ik} + B^{(g)}_{kj}\bigr) - 2kj + \bigl(j^{2} - \lambda v_{j} - F[j]\bigr).
\]
Let $T^{*}_{ij} = (A^{(g)} \otimes B^{(g)})_{ij} \in \mathbb{Z}$. For any minimizing index $k^{*}$, we have $0 \le 2k^{*}j \le 2(n-1)^{2} < \lambda = 2n^{2} + 1$, so $\lambda T^{*}_{ij} - 2k^{*}j \in [\lambda T^{*}_{ij} - 2(n-1)^{2}, \lambda T^{*}_{ij}]$. For any $k$ with $A^{(g)}_{ik} + B^{(g)}_{kj} \ge T^{*}_{ij} + 1$, we have $\lambda(A^{(g)}_{ik} + B^{(g)}_{kj}) - 2kj \ge \lambda T^{*}_{ij} + \lambda - 2(n-1)^{2} > \lambda T^{*}_{ij}$. Hence $\widetilde{S}_{ij} = S_{ij} - j^{2} + \lambda v_{j} + F[j]$ satisfies $\lambda(T^{*}_{ij} - 1) < \widetilde{S}_{ij} \le \lambda T^{*}_{ij}$, so $\lceil \widetilde{S}_{ij} / \lambda \rceil = T^{*}_{ij}$ exactly. All constructed weights satisfy $|D_{pq}| \le 4U$ and $|F[j]| = \mathcal{O}(n^{4}U)$, preserving $\mathcal{O}(\log n)$-bit word size. Constructing all $Q = \lceil n/b \rceil = \mathcal{O}(n^{1-\alpha})$ instances takes $\mathcal{O}(n^{3-\alpha})$ total time, and invoking the $\mathcal{O}(n^{\beta})$ solver $Q$ times takes $\mathcal{O}(n^{1-\alpha+\beta})$ time.
\end{proof}

\begin{algorithmsteps}{gap-cuts}{Matrix product by vertex cover cuts}
1: & Greedily match unmatched $(a,b)$ with $\Delta B_{a,b} < 0$ and select both $a$ and $b$ as cut gaps \\*
2: & Partition rows into $R_{1}, \dots, R_{p}$ and columns into $J_{1}, \dots, J_{q}$ ($p + q \le 2\tau(B) + 2$) \\*
3: & \textbf{for} $i = 0, \dots, n-1$ \textbf{do} initialize $C_{i,0..n-1} \leftarrow +\infty$; \textbf{for} each block $R_{a} \times J_{b}$ \textbf{do} use SMAWK on $(A_{ik} + B_{kj})_{k \in R_{a}, j \in J_{b}}$ to update $C_{i, j \in J_{b}}$ \textbf{end for end for} \\
\end{algorithmsteps}

\begin{proof}[Proof of Theorem~\ref{thm:support-graph}]
\textbf{Vertex cover.} In Algorithm~\ref{alg:gap-cuts}, a greedy maximal matching of $G_{-}(B)$ finds at most $\tau(B)$ disjoint edges in $\mathcal{O}(n^{2})$ time, and taking both endpoints of each matched edge yields a vertex cover of size at most $2\tau(B)$, partitioning rows and columns into $p + q \le 2\tau(B) + 2$ intervals. By telescoping~\eqref{eq:telescoping}, every block $R_{a} \times J_{b}$ contains no negative cell in its interior and is therefore Monge, and adding $A_{ik}$ preserves Monge structure. Applying SMAWK~\cite{Aggarwal1987} to each block costs $\mathcal{O}(|R_{a}| + |J_{b}|)$ per row $i$, summing to $\sum_{a=1}^{p}\sum_{b=1}^{q} (|R_{a}| + |J_{b}|) = nq + np = \mathcal{O}(n(\tau(B)+1))$ per row and $\mathcal{O}(n^{2}(\tau(B)+1))$ overall.

\textbf{Lower bounds.} For $s,\tau,t$, the padded group matrices in the proof of Theorem~\ref{thm:fixed-quad-lb} give $B^{(g)}$ with $s(B^{(g)}) \le b(n-1) \le n^{1+\alpha}$, $\tau(B^{(g)}) \le b \le n^{\alpha}$, and pathwidth (hence treewidth) $t(B^{(g)}) \le b \le n^{\alpha}$ using bags $R \cup \{c\}$ where $R$ is the set of active row gaps. For maximum degree $\Delta_{-}$, partition rows and columns into $Q = \lceil n/b \rceil$ blocks $I_{0}, \dots, I_{Q-1}$ and $J_{0}, \dots, J_{Q-1}$ of size at most $b = \lfloor n^{\alpha}/4 \rfloor$. For $g = 0, \dots, Q-1$, set $B^{(g)}[I_{a}, J_{c}] = B[I_{a}, J_{c}]$ if $c \equiv a + g \pmod{Q}$ and $2U + 1$ otherwise. Every entry of $B$ appears in exactly one $B^{(g)}$, and every row or column gap touches at most two retained blocks of width $\le b$, giving $\Delta_{-}(B^{(g)}) \le 2b + 2 \le n^{\alpha}$ for sufficiently large $n$. Taking the entrywise minimum of $A \otimes B^{(g)}$ over all $Q = \mathcal{O}(n^{1-\alpha})$ cyclic shifts recovers $A \otimes B$ in $\mathcal{O}(n^{1-\alpha+\beta} + n^{3-\alpha})$ time.
\end{proof}

\besidetext{.43\linewidth}{%
    \noindent\algheading{repeated-mult}\enspace Repeated matrix products\par
    \smallskip
    \begin{tabularx}{\linewidth}{@{}r@{\enspace}>{\raggedright\arraybackslash}X@{}}
        \toprule
        1. & Set $X\gets X_0$. \\
        2. & For each $t=1,\ldots,\ell$, run Algorithm~\ref{alg:explicit-product} on $X,B_t$. \\
        3. & At each step replace $X$ by the product and retain $W_t$ if a path is requested. \\
        4. & Return $X$ and the retained witnesses. \\
        \bottomrule
    \end{tabularx}\par
}{%
    For successive right factors, Algorithm~\ref{alg:repeated-mult} uses the current product as the left factor, so at layer $t$ the running time depends on the negative support of $B_t$. To recover a path from source $i$ to terminal $j$, one starts in column $j$ of the last layer and reads the stored witnesses in reverse order, keeping the source fixed and using each intermediate index as the column in the preceding layer.
}
When only the final distances are requested, successive products reuse the same working storage.

\subsection{Product values from Russo's implicit Monge product}
\label{app:russo-alt}

For Monge matrices $U,V$ with constant-time entry access, Russo's product representation~\cite[Lemma~6]{Russo2010} takes $\mathcal{O}(nL+\delta(U)L^3)$ construction time and $\mathcal{O}(n+\delta(U)L)$ space, where $\delta(U)$ is the number of nonzero adjacent differences of $U$, and answers an entry query to $U\otimes V$ in $\mathcal{O}(L)$ time.

With $w_{p,q}=\max\{-\Delta B_{p,q},0\}$, the prefix sums $P(k,j)=\sum_{p<k,\,q<j}w_{p,q}$ satisfy $\Delta P_{p,q}=-w_{p,q}$, so $M=B-P$ is Monge. For a row $x=A_{i,*}$, the matrix $U(a,k)=x_k+P(k,n-1-a)$ for $0\le a,k<n$ has differences $\Delta U_{a,k}=w_{k,n-2-a}\ge0$ and hence is Monge with core size $\delta(U)=s(B)$, while the entries on the antidiagonal of $V=U\otimes M$ satisfy
\[
V(n-1-j, j) = \min_{0 \le k < n} \{x_{k} + P(k, j) + M(k, j)\} = (x \otimes B)_{j}.
\]
Querying the $n$ antidiagonal entries for each row computes $A\otimes B$ in $\mathcal{O}(n^2L+nsL^3)$ time and $\mathcal{O}(n^2+sL)$ space using the implicit products, and the envelope sweep of Algorithm~\ref{alg:explicit-product} computes the values and witnesses in the same time with $\mathcal{O}(n^2)$ space.

\section{Further questions}
\label{app:open}

\begin{itemize}
    \item \textbf{Updates for internal rewards.} The three logarithmic factors in $\mathcal{O}((n+r)L^3)$ come from visiting the ancestors at the boundary of an interval, searching for a switch at each such ancestor, and evaluating an envelope at each search step. For an internal reward at $j=\max e$, the affected rows satisfy $k<\min e\le j$, so all earlier events miss row $j-1$ and the sweep can activate it with $D[j-1]$ before processing the current bucket, including singleton interactions, using the envelope tree alone.
    \item \textbf{Finding rectangle representations.} For a cost array supplied by an entry oracle, open problems include determining the number of queries needed to locate $\mathcal{N}(C)$ and finding a representation with the smallest number of signed rectangles, accounting for cancellation where rectangles overlap.
    \item \textbf{Vertex order.} The partition algorithms take the vertex order as input, and the effect of using a spectral order or an order with small total edge length on the optimum and the running time remains to be determined for sparse graphs and hypergraphs.
    \item \textbf{Vector queries.} Corollary~\ref{cor:adaptive-query-lower} assumes quadratic preprocessing and arbitrary query vectors, leaving open the query exponents achievable with a larger preprocessing budget or with a prescribed relation between successive vectors.
\end{itemize}

\section*{AI assistance}
The author used GPT-6-Astra and GPT-6.1-Sol to assist with literature searches and proofs for an earlier version and takes responsibility for the content of the paper.

\end{document}